\documentclass{aa}  

\usepackage{graphicx}
\usepackage{setspace}
\usepackage{natbib}
\usepackage{rotating}
\usepackage{threeparttable}
\usepackage{tabto}
\usepackage{float}
\usepackage{subcaption}
\usepackage{amsmath}
\usepackage{nccmath}
\usepackage{orcidlink}

\bibpunct{(}{)}{;}{a}{}{,} 

\usepackage{txfonts}
\usepackage{hyperref}
\hypersetup{
     colorlinks=true,
     linkcolor=blue,
     filecolor=blue,
     citecolor = blue,      
     urlcolor= blue,
     }
\begin{document} 

   \title{Investigating episodic mass loss in evolved massive stars:}
   
    \subtitle{II. Physical properties of red supergiants at subsolar metallicity\thanks{Based on observations collected at the European Southern Observatory under ESO programs 105.20HJ and 109.22W2.}}




   \author{S. de Wit \inst{1,2}\orcidlink{0000-0002-9818-4877} \and 
          A.Z. Bonanos \inst{1}\orcidlink{0000-0003-2851-1905} \and 
          K. Antoniadis \inst{1,2}\orcidlink{0000-0002-3454-7958} \and 
          E. Zapartas \inst{3,1} \orcidlink{0000-0002-7464-498X} \and
          A. Ruiz \inst{1}\orcidlink{0000-0002-3352-4383} \and \\
          N. Britavskiy \inst{4,5}\orcidlink{0000-0003-3996-0175} \and
          E. Christodoulou \inst{1,2}\orcidlink{0000-0003-4332-3646} \and
          K. De \inst{6}\orcidlink{0000-0002-8989-0542} \and
          G. Maravelias \inst{1,3}\orcidlink{0000-0002-0891-7564} \and \\
          G. Munoz-Sanchez \inst{1,2}\orcidlink{0000-0002-9179-6918} \and 
          A. Tsopela \inst{7} \orcidlink{0009-0008-1732-3638}
          }

   \institute{
    IAASARS, National Observatory of Athens, 15326 Penteli, Greece
    \and
    National and Kapodistrian University of Athens, Department of Physics, Panepistimiopolis, 15784 Zografos, Greece 
    \and
    Institute of Astrophysics FORTH, 71110 Heraklion, Greece
    \and
    University of Li\`ege, All\'ee du 6 Ao\^ut 19c (B5C), B-4000 Sart Tilman, Li\`ege, Belgium
    \and Royal Observatory of Belgium, Avenue Circulaire/Ringlaan 3, B-1180 Brussels, Belgium
    \and
    Kavli Institute for Astrophysics and Space Research, Massachusetts Institute of Technology, Cambridge, MA, USA  
    \and
    Department of Physics and Astronomy, Dartmouth College, 6127 Wilder Laboratory, Hanover, NH 03755, USA
    }    
   \date{}
   
 
  \abstract
  {Mass loss during the red supergiant (RSG) phase plays a crucial role in the evolution of an intermediate massive star, however, the underlying mechanism remains unknown. We aim to increase the sample of well-characterized RSGs at subsolar metallicity, by deriving the physical properties of 127 RSGs in nine nearby, southern galaxies presented by Bonanos et al. For each RSG, we provide spectral types and used \textsc{marcs} atmospheric models to measure stellar properties from their optical spectra, such as the effective temperature, extinction, and radial velocity. By fitting the spectral energy distribution, we obtained the stellar luminosity and radius for 97 RSGs, finding $\sim 50\%$ with log$(L/ \rm L_{\odot}) \geq 5.0$ and 6 RSGs with $R \gtrsim 1400 \,\ \rm R_{\odot}$. We also find a correlation between the stellar luminosity and mid-IR excess of 33 dusty, variable sources. Three of these dusty RSGs have luminosities exceeding the revised Humphreys-Davidson limit. We then derive a metallicity-dependent $J-K_s$ color versus temperature relation from synthetic photometry and two new empirical $J-K_s$ color versus temperature relations calibrated on literature TiO and $J$-band temperatures. To scale our derived, cool TiO temperatures to values in agreement with the evolutionary tracks, we derive two linear scaling relations calibrated on $J$-band and $i$-band temperatures. We find that the TiO temperatures are more discrepant as a function of the mass-loss rate and discuss future prospects of the TiO bands as a mass-loss probe. Finally, we speculate that 3 hot, dusty RSGs may have experienced a recent mass ejection ($12\%$ of the K-type sample) and indicate them as candidate Levesque-Massey variables.}


  \titlerunning{Physical properties of red supergiants at subsolar metallicity}
   \authorrunning{de Wit et al.}

\maketitle


\section{Introduction}

Massive stars significantly contribute to the evolution of galaxies by providing mechanical feedback and chemically enriching the interstellar medium. It is, therefore, crucial to understand their evolution and feedback mechanism, specifically those channeled through stellar winds and outbursts, which may significantly alter the late-stage evolution of a massive star, as well as the properties of the resulting supernova and compact remnant \citep{Smith2014}. The red supergiant (RSG) phase is a significant contributor \citep{Levesque2017, Decin2021} to these mass-loss channels, as the majority of massive stars either evolve through this phase, either transitioning into yellow supergiants or yellow hypergiants \citep[e.g.,][]{Jager1998, Gordon2019}, or terminate their lives as a RSG via a core-collapse supernova or direct implosion to a black hole \citep{Heger2003, Sukhbold2016, Laplace2020}. Despite losing mass at higher rates in the RSG phase compared to their OB main-sequence phase \citep[$\sim$~1 order of magnitude, depending on the metallicity;][]{Beasor2021}, RSG outflows are still subject to many uncertainties, on both theoretical and empirical grounds \citep[see e.g.,][]{Antoniadis2024}. Adding to the already complex stellar outflows are clues for episodic mass loss \citep{Smith2014, Bruch2021, Humphreys2022}.

The mass-loss behavior of RSGs, although significant, remains elusive, despite many recent efforts to study the mass-loss rates and driving mechanisms observationally and theoretically \citep[e.g.,][]{Beasor2020, Kee2021, Davies2021, Decin2023}. \cite{Kee2021} have suggested that turbulent motions provide the force to steadily drive material off the stellar surface, and as such, the escape mechanism likely does not depend on the radiation field \citep[although recently challenged by][]{Vink2023}. The proposed mass-loss mechanism(s) should be time-dependent, yet we often adopt instantaneous mass-loss rates \citep{Jager1988, Loon2005, Goldman2017, Beasor2020, Yang2023, Antoniadis2024} to characterize mass loss throughout the RSG phase. \citet{Massey2022}, however, used the luminosity function to study the role of a time-averaged mass-loss rate. 

Empirical prescriptions may differ by 2-3 orders of magnitude (e.g., \citealt{Jager1988} versus \citealt{Goldman2017}). Adopting one or the other drastically impacts the post-main-sequence evolution, stripping the hydrogen-rich envelope, which then affects the supernova type and compact remnant properties after core-collapse \citep{Laplace2020, Laplace2021} when implemented into stellar evolution models \citep[e.g. \textsc{mesa};][]{Paxton2011, Paxton2013, Paxton2015, Paxton2018}. Currently, some empirical prescriptions \citep{Kee2021, Yang2023} consistently strip RSG envelopes down to $M_{\rm ini} \sim 15 \rm M_{\odot}$ (Zapartas et al., in prep.), while some other prescriptions fail to do so \citep{Beasor2020, Decin2023}, undermining observations (i.e. the lack of exploding RSGs of $M_{\rm ini} \geq 20 \rm M_{\odot}$; the ``RSG problem'', see e.g., \citealt{Smartt2009, Davies2020, Beasor2021}, or the existence of post-RSG objects, such as "F15" in \citealt{Decin2023}). Therefore, such observations are typically explained by interactions in binary systems \citep{Eldridge2013, Zapartas2017} or short-lived pre-supernova explosions \citep{Smith2009}. 

Adding to the complexity of the stellar wind is the variety of assumptions for deriving $\dot M$ (e.g., gas-to-dust ratio, geometrical symmetries, dust composition, and grain size distribution), as well as uncertainties on the surface properties forming at the base of the wind. An accurately constrained RSG effective temperature ($T_{\rm eff}$) is specifically important for spectral energy distribution (SED) fitting models to obtain mass-loss rates \citep[e.g., \textsc{dusty};][]{Dusty1}. Furthermore, $T_{\rm eff}$ is a direct input into several common mass-loss prescriptions \citep[e.g.,][]{Jager1988,Nieuwenhuijzen1990,Loon2005}. Not only does $T_{\rm eff}$ strongly depend on the metallicity, it is also variable due to intrinsic velocity cycles \citep[hysteresis loops;][]{Kravchenko2019, Kravchenko2021}. Moreover, changes in the immediate circumstellar environment due to high mass loss \citep[log($\dot M) \leq -5.0~\rm{dex}$;][]{Davies2021} may significantly hinder the derivation of $T_{\rm eff}$. Both the hysteresis loops and mass loss change the appearance of the emerging spectrum, as the location of the photosphere moves in or outwards \citep[see e.g.,][]{Massey2007, Levesque2007}. To understand stellar outflows, it is crucial to understand the evolutionary state of the RSG (e.g. $T_{\rm eff}$, $L_{\star}$, which yield a location in the Hertzsprung-Russell diagram; HRD) and the force required to lift material off the surface (e.g., the escape velocity; $v_{\rm esc}$ from the surface gravity; $\rm log\,\it g$ and the turbulent velocity; $v_{\rm turb}$). Not only do spectral properties of RSGs change but it has also been established that RSGs are photometric semi-regular variables, varying up to a few magnitudes in the optical \citep{Kiss2006}. Luminous RSGs even display significant variability in their infrared (IR) magnitudes \citep[e.g., WISE;][]{Yang2018}.

Many different methods exist to obtain the surface properties of RSGs, depending on the wavelength domain in which they were observed. Nearby RSGs, such as Betelgeuse, are resolved and allow for the extraction of complex surface characteristics, such as velocity variations and shifting temperatures at the stellar surface \citep[either due to rapid rotation or convective motions;][]{Kervella2018, Ma2023}. For extragalactic RSGs, many studies have targeted nearby galaxies (Local Group, i.e. LMC, SMC, NGC~6822, WLM and Sextans A, but also up to 2~Mpc, i.e. NGC~300 and NGC~55) with spectroscopy, spanning a range of metallicities, to obtain the effective temperature: \cite{Levesque2005, Levesque2006} fitted titanium-oxide (TiO) molecules in the optical, \cite{Davies2013} fitted the line-free infrared continuum, \cite{Patrick2015, Patrick2017} and \cite{Gazak2015} fitted spectral lines in the $J$-band, \cite{Tabernero2018} fitted spectral lines in the $i$-band and lastly, \cite{Gonzalez2021} fitted absorption-free spectral windows throughout the whole spectral energy distribution (SED). Most of the methods applied in these studies lead to temperatures that are in disagreement. This is likely explained by the stellar atmosphere models used, which are typically one-dimensional \citep[e.g., \textsc{marcs};][]{Gustafsson2008}, yet the atmospheres of RSGs have a much more complex three-dimensional structure and motions \citep[e.g. convection and granulation, see][]{Freytag2002, Ludwig2009, Chiavassa2011}. 

The temperatures obtained through the TiO diagnostic are the main outlier \citep{Davies2013}, as they systematically yield a lower $T_{\rm eff}$ for a given RSG. This discrepancy has been attributed to increased molecular absorption in the \textsc{marcs} models, potentially due to a combination of two effects. Firstly, the 1D assumptions in the \textsc{marcs} models do not precisely represent a 3D RSG atmosphere \citep{Davies2013}, specifically regarding the temperature gradient in the stellar atmosphere. Secondly, the \textsc{marcs} models ignore the effects of mass loss on the spectral appearance \citep{Davies2021}, which significantly alters the TiO opacity. Furthermore, \cite{Davies2013} demonstrated that the $A_V$ values derived from TiO-dominated spectra are too low, compared to existing extinction maps and the strengths of diffuse interstellar bands of the same spectrum. As such, these authors urged caution when interpreting $T_{\rm eff}$ and $A_V$ derived from the TiO diagnostic, given the current state of the \textsc{marcs} models. However, spectroscopic observations are typically obtained in the optical due to the wide availability of optical spectrographs, particularly for multi-object spectroscopy. Therefore, there is an increasing need for a new $T_{\rm eff}$ scaling relation to convert between TiO temperatures and the more reliable $J$-band or $i$-band temperatures. 

A large catalog of RSGs beyond the Local Group has recently been presented by the ASSESS\footnote{The ASSESS website: \url{http://assess.astro.noa.gr/}} project, resulting from optical spectroscopy, aiming to study the role of episodic mass loss in evolved massive stars \citep{Bonanos2023VLT}. The next step is to characterize the physical properties of these 127 RSGs with spectral modeling. In Sect.~\ref{SecData}, we briefly summarize the observations and target selection, provide an updated spectral classification for each RSG, obtain a stellar luminosity for most RSGs from SED-fitting, and present RSG light curves from various surveys. In Sect.~\ref{SecMeth}, we present the spectral modeling strategy and the grid of evolutionary models used in this work. In Sect.~\ref{SecRes}, we present the stellar properties of our sample of RSGs and derive scaling relations to re-interpret the effective temperatures obtained in this work. We discuss our results in Sect.~\ref{SecDisc}. We summarize the results in Sect.~\ref{SecSum}.

\section{Data} \label{SecData}
\subsection{Target selection and observations} \label{Data:Selection}

\begin{figure*}[h]
\begin{center}
    \centerline{\includegraphics[width=2\columnwidth]{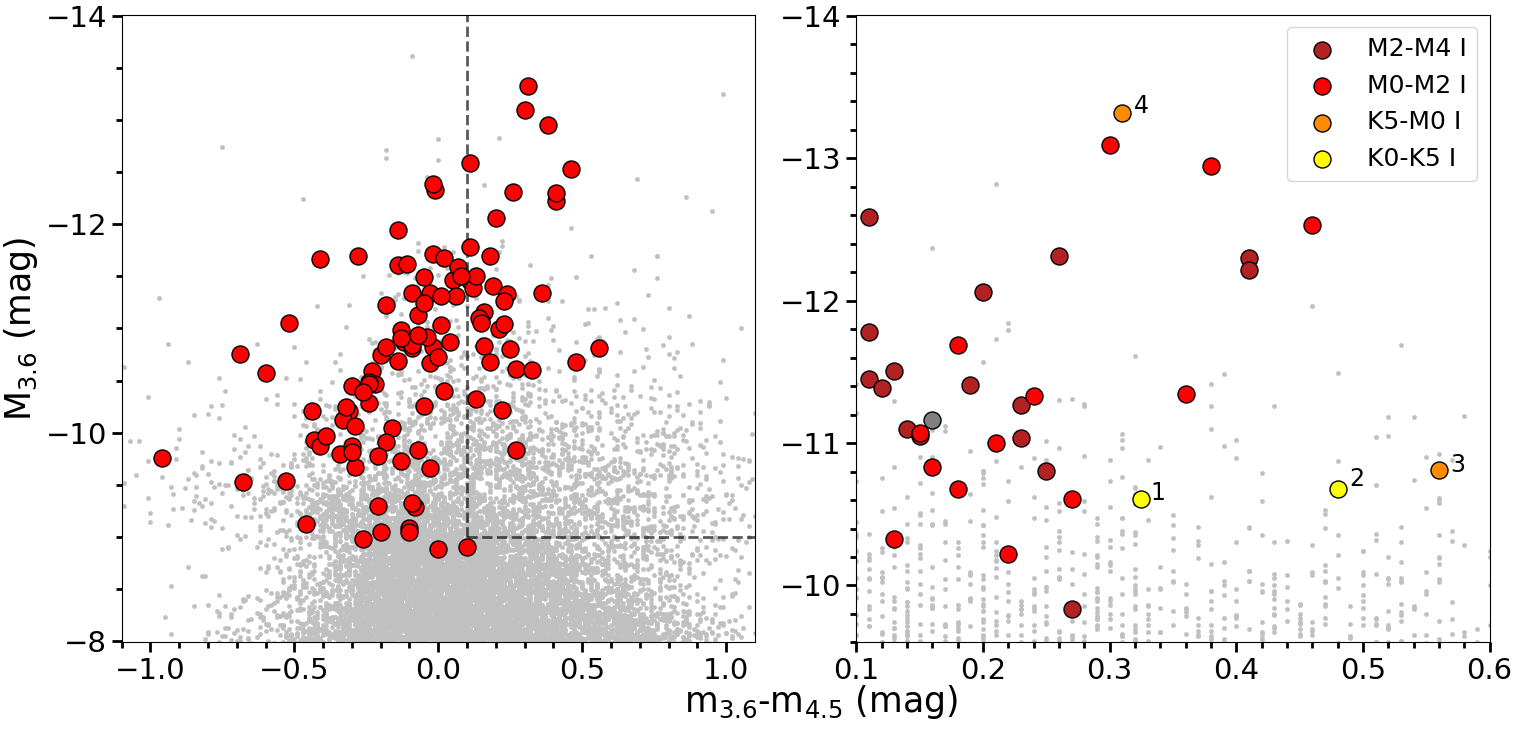}} 
    \caption{\textit{Left}: The [3.6]$-$[4.5] versus M$_{[3.6]}$ of 129 studied RSGs. Gray background points are sources from the \textit{Spitzer} point source catalog of NGC~300. The dashed lines indicate the region of prioritized sources described in \cite{Bonanos2023VLT}, showing some degree of IR excess. \textit{Right}: Zoom in on 34 prioritized, dusty targets (33 classified). Red to yellow colors indicate a shift in spectral type towards hotter RSGs. The gray circle indicates the unclassified RSG. Four dusty K-type RSGs are highlighted: NGC3109-167~(1), NGC55-244~(2), NGC55-202~(3) and M83-682~(4).}
    \label{AbsCMD}
\end{center}
\end{figure*}

\begin{table*}
\centering     
  \begin{threeparttable}
    \caption{Properties of galaxies studied in this work.}
    \tiny
    \label{GalaxyTable}
    \renewcommand{\arraystretch}{1.2}
        \begin{tabular}{l l l l l l r r}        
        \hline\hline                 
        Galaxy & RA & Dec & Distance & $Z$ & $v_{\rm rad}$ & \# M-RSG & \# K-RSG \\ 
         & \small{(J2000)}	& \small{(J2000)} & (Mpc) & (Z$_{\odot}$) & (km s$^{-1}$) &   &  \\    
        \hline\hline                        
            NGC~55 & 00:14:53.60    & $-$39:11:47.9 & 1.87$\pm$0.02$^1$    & 0.27$^2$    & 129$\pm$2$^3$ & 27 & 15\\
            NGC~247 & 00:47:08.55	& $-$20:45:37.4 & 3.03$\pm$0.03$^1$    & 0.40$^4$    & 156$^5$       & 14 & 2\\
            NGC~253 & 00:47:33.12 	& $-$25:17:17.6 & 3.40$\pm$0.06$^6$ & 0.72$^7$ & 259$^8$    & 11 & 1\\
            NGC~300 & 00:54:53.48	& $-$37:41:03.8 & 1.97$\pm$0.06$^1$    & 0.41$^9$    & 146$\pm$2$^3$ & 39 & 7\\ 
            NGC~1313 & 03:18:16.05	& $-$66:29:53.7 & 4.61$\pm$0.17$^{10}$ & 0.35$^{11}$ & 470$^{12}$    & 0  & 1\\
            NGC~3109 & 10:03:06.88	& $-$26:09:34.5 & 1.27$\pm$0.03$^1$    & 0.21$^{13}$    & 403$\pm$2$^3$ & 0  & 1\\
            Sextans~A & 10:11:00.80	& $-$04:41:34.0 & 1.34$\pm$0.02$^{14}$    & 0.06$^{15}$    & 324$\pm$2$^3$ & 0  & 1\\
            M83      & 13:37:00.95	& $-$29:51:55.5 & 4.90$\pm$0.20$^{16}$ & 1.58$^{17}$ & 519$^8$    & 0  & 1 \\
            NGC~7793 & 23:57:49.83	& $-$32:35:27.7 & 3.47$\pm$0.04$^1$    & 0.42$^{18}$ & 227$^8$    & 8  & 1\\
        \hline
        \end{tabular}
    \begin{tablenotes}
        \small
        \vspace{5pt}
        \item References: $^1$\cite{Zgirski2021}, $^2$\cite{Hartoog2012}, $^3$\cite{McConnachie2012}, $^4$\cite{Davidge2021}, $^5$\cite{Tully2016}, $^6$\cite{Madore2020}, $^7$\cite{Spinoglio2022}, $^8$\cite{Meyer2004}, $^9$\cite{Kudritzki2008}, $^{10}$\cite{Qing2015}, $^{11}$\cite{Hadfield2007}, $^{12}$\cite{Koribalski2004}, $^{13}$\cite{Hosek2014}, $^{14}$\cite{Tammann2011}, $^{15}$\cite{Kniazev2005}, $^{16}$\cite{Bresolin2016}, $^{17}$\cite{Hernandez2019}, $^{18}$\cite{DellaBruna2021}.
    \end{tablenotes}
    \end{threeparttable}
\end{table*}

\cite{Bonanos2023VLT} presented 129 RSGs located in nine southern galaxies up to 5~Mpc, which were selected primarily by their mid-IR magnitudes and colors. In Table~\ref{GalaxyTable}, we list the targeted galaxies ordered by right ascension (RA), and their properties such as distance, metallicity, radial velocity, and the number of M-type and K-type RSGs. For 117 of these, no matches were found with spectroscopically classified objects in the literature and were therefore viewed as new discoveries. The other 12 RSGs have counterparts in the literature (five were identified as `RSG' in \citealt{Gazak2015}, six were identified as `RSG' in \citealt{Patrick2017} and one was classified as G2~I in \citealt{Evans2007}). \cite{Bonanos2023VLT} imposed strict criteria on the IRAC [3.6] and [4.5] photometric bands, to select dusty, evolved massive stars. These bands were chosen to estimate the amount of IR excess (i.e. due to a hot dust-rich component), which may indicate past episodes of enhanced mass loss.

Here, we break down the criteria as follows: \textit{(i)} M$_{3.6}\le-9.0$~mag; to avoid contamination from the asymptotic giant branch (AGB) stars \citep{Yang2020}, \textit{(ii)} m$_{4.5}\le15.5$~mag; to avoid contamination by background galaxies \citep{Williams2015} and {(\textit{iii})} $\textrm{m}_{3.6}-\textrm{m}_{4.5}>0.1$~mag; to avoid contamination by foreground stars. When a source passed all of these criteria, it was observed as a priority target ($f_{prio} \sim 26\%$), increasing in priority as the IR excess increased (P1; $\textrm{m}_{3.6}-\textrm{m}_{4.5}>0.5$~mag). A source was observed as a non-priority target ("filler") if it satisfied none of these criteria. For the RSG sample, 79$\%$ are fillers and the vast majority of these do not show IR excess. 

We show the location of our RSGs in the \textit{Spitzer} color-magnitude diagram (CMD) in the left panel Fig.~\ref{AbsCMD}. Dusty sources with IR excess ($\textrm{m}_{3.6}-\textrm{m}_{4.5}>0.1$~mag) are shown in the right panel of Fig.~\ref{AbsCMD}. We note that the IR excess is not necessarily explained by dust and may also arise from molecular emission of SiO and CO bands in the mid-IR \citep{Verhoelst2009, Davies2021} by a RSG with a high mass-loss rate. Contrarily, some of the blue RSGs in Fig.~\ref{AbsCMD} ($\textrm{m}_{3.6}-\textrm{m}_{4.5}<-0.5$~mag) may be dusty due to PAH emission around $\sim 3.3 \mu$m, as speculated in \cite{Yang2018}. To convert the apparent magnitudes for each source to an absolute magnitude, we used the distances from Table~\ref{GalaxyTable}. The target IDs, coordinates, magnitudes, spectral types (see Sect.~\ref{Data:Class}), variability properties, and light curves (see Sect.~\ref{Data:LCs}) are presented in Table~\ref{PhotoTable}. This table contains identical data to those presented in \cite{Bonanos2023VLT}, except for the \textit{Gaia}~DR2 magnitudes, which have been replaced with \textit{Gaia}~DR3 magnitudes (Table~\ref{PhotoTable}), and the updated distances to the galaxies in Table~\ref{GalaxyTable}.

Following our target selection strategy, apart from the RSGs studied here, we observed a wide variety of evolved massive stars, such as luminous blue variable candidates, yellow supergiants, and supergiant B[e] stars, and provided an initial classification in both \cite{Bonanos2023VLT} and \cite{Maravelias2023}. The spectra were taken by the VLT using the multi-object spectroscopy mode of the Focal Reducer and Low Dispersion Spectrograph (FORS2) under ESO programs 105.20HJ and 109.22W2. The observing details, grism choice, and reduction process were presented in \cite{Bonanos2023VLT}. The resulting spectra yielded a resolving power of $R\sim$1000 and provided a wavelength coverage from $\sim$5300-8450~\AA. A direct determination of the luminosity class was not possible, as the Ca~\textsc{ii} triplet ($\sim$8500-8700~\AA) was not covered.

\begin{table*}
\centering     
  \begin{threeparttable}
    \caption{Photometric properties of our RSG sample.}
    \tiny
    \label{PhotoTable}
    \renewcommand{\arraystretch}{1.3}
        \begin{tabular}{l l l c c c l c c l l l}        
        \hline\hline                 
        ID	&	RA & Dec &	G	&	BP	&	RP & ... & [3.6]	&	[4.5]	&	Spectral &	 MAD$_{\rm W1}$	& Light curve\\
         & (J2000 deg) & (J2000 deg)	&	(mag)	&	(mag) & (mag)	&	... &	(mag)	&	(mag)	& class &	& bands	 \\
        \hline\hline                        
NGC55-40 &	3.69121	& $-$39.17961	&	18.687	&	20.690	&	17.574	&	...	&	14.14	&	13.73	&	M2-M4~I	&	0.016	&	ATLAS-c,o	\\
NGC55-75 &	3.87163	& $-$39.23562	&	18.935	&	20.011	&	17.893	&	...	&	14.86	&	14.78	&	K5-M0~I	&	0.043	&	ATLAS-c,o	\\
NGC55-87 &	3.88813	& $-$39.22835	&	19.458	&	20.628	&	18.315	&	...	&	14.95	&	14.76	&	M2-M4~I	&	0.034	&		\\
NGC55-93 &	3.79834	& $-$39.21316	&	19.793	&	20.268	&	18.420	&	...	&	14.97	&	14.85	&	M2-M4~I	&	0.020	&		\\
NGC55-135 &	3.60979	& $-$39.16807 &	19.036	&	20.133	&	17.975	&	...	&	15.26	&	15.12	&	M2-M4~I	&	0.026	&	ATLAS-c,o	\\
NGC55-146 &	3.91958	& $-$39.24735	&	19.811	&	20.284	&	18.420	&	...	&	15.31	&	15.16	&	M2-M4~I	&	0.012	&		\\
NGC55-147 &	3.73659	& $-$39.20215	&	20.996	&	22.509	&	19.860	&	...	&	15.32	&	15.09	&	M2-M4~I	&	0.006	&		\\
NGC55-149 &	3.63609	& $-$39.17836	&	20.411	&	20.761	&	19.082	&	...	&	15.36	&	15.15	&	M0-M2~I	&	0.013	&		\\
NGC55-152 &	3.65251	& $-$39.17726	&	19.885	&	20.453	&	18.429	&	...	&	15.37	&	15.50	&	M0-M2~I	&	0.011	&		\\
NGC55-165 &	3.89187	& $-$39.22507	&	18.670	&	19.568	&	17.722	&	...	&	15.42	&	15.49	&	K0-K5~I &	0.019	&	ATLAS-c,o \\
NGC55-174 &	3.79936	& $-$39.20461	&	19.570	&	20.398	&	18.451	&	...	&	15.44	&	15.48	&	M0-M2~I	&	0.018	&		\\
NGC55-194 &	3.84200	& $-$39.22041	&	20.039	&	20.636	&	18.742	&	...	&	15.52	&	15.61	&	M0-M2~I	&	0.012	&		\\
NGC55-200 &	3.98479	& $-$39.26915	&	20.107	&	21.279	&	19.089	&	...	&	15.54	&	15.56	&	M2-M4~I	&	0.020	&		\\
NGC55-202 &	3.70461	& $-$39.20299 &	20.251	&	21.370	&	19.676	&	...	&	15.55	&	14.99	&	K5-M0~I	&	0.012	&		\\
NGC55-216 &	3.81259	& $-$39.22191	&	19.449	&	20.011	&	18.375	&	...	&	15.61	&	15.81	&	M0-M2~I	&	0.009	&		\\
        \vdots	&	\vdots	 & \vdots	&	\vdots	&	\vdots	&	\vdots	&	\vdots	&	\vdots	&	\vdots	&	\vdots	&	\vdots \\
        \hline
        \end{tabular}
    \begin{tablenotes}
        \small
        \vspace{5pt}
        \item Notes: This table is available in its entirety in machine-readable and Virtual Observatory (VO) forms in the online journal. A portion is shown here for guidance regarding its form and content.
    \end{tablenotes}
    \end{threeparttable}
\end{table*}

\subsection{Spectral classification}\label{Data:Class}

The RSGs presented in \citet{Bonanos2023VLT} were initially broadly classified into spectral class K or M. Here, we revisited the spectra and assigned an approximate spectral type (e.g., K0-K5, K5-M0, M0-M2, and M2-M4). It was not possible to determine the exact spectral type, due to the low spectral resolving power, low signal-to-noise ratio, or missing diagnostics (i.e. the Ca~\textsc{ii} triplet and, in nine cases, both sets of Na~\textsc{ii} and K~\textsc{i} lines). When the TiO bands were either weak or completely absent, we assigned a K0-K5 spectral type. If the TiO bands at $\lambda\lambda$6150 and $\lambda\lambda$7050 were somewhat present, we assigned a K5-M0 spectral type. We assigned a spectral type M0-M2 to spectra displaying moderately strong $\lambda\lambda$6150 and $\lambda\lambda$7050 bands, but no molecular TiO absorption at $\lambda\lambda$7600. Finally, if all TiO bands were strong, and the spectrum showed significant TiO absorption at $\lambda\lambda$7600, we assigned an M2-M4 spectral type. We were unable to determine a spectral type for two RSGs (NGC300-237 and NGC300-609), due to spectral reduction artifacts. The revised spectral types of the RSGs are listed under "Spectral class" in Table~\ref{PhotoTable}.

\begin{figure}[h]
\begin{center}
\centering
    \includegraphics[width=1\columnwidth]{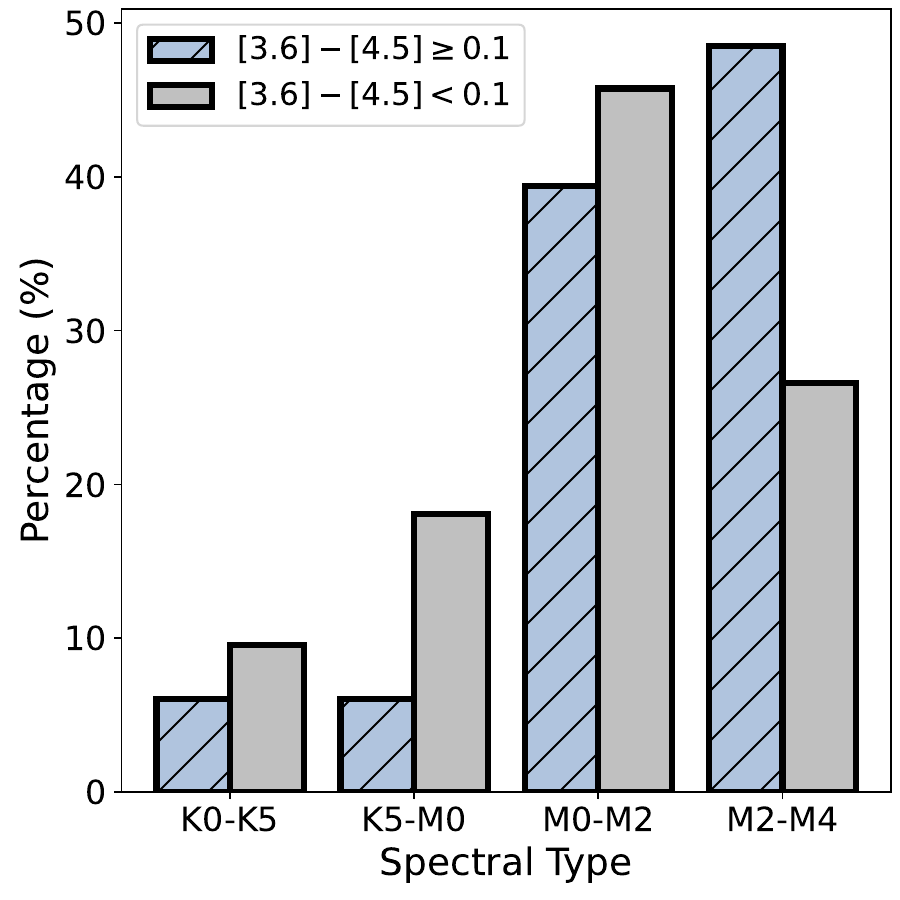}
    \caption{Distribution of spectral types of classified RSGs. Blue hatched bars show RSGs with IR excess (see right panel of Fig.~\ref{AbsCMD}; 33 targets). The remaining RSGs (94) are shown with gray bars.}
    \label{fig:SpTDist}
\end{center}
\end{figure}

Despite the spectra showing only features of single RSGs, we cannot exclude two possible groups of contaminants. Firstly, given their intrinsic brightness and similarity in spectral features, luminous extragalactic oxygen-rich AGB stars may contaminate our sample \citep[although extremely rare, they can reach up to log~$(L/L_{\odot}) \sim 5.0$;][]{Groenewegen2018}. Secondly, especially for more distant galaxies, the observed spectrum may be a composite of a RSG and other stars which do not significantly contribute to the optical spectrum but do affect the photometry presented in Fig.\ref{AbsCMD} and Table~\ref{PhotoTable}. 

We show the distribution of spectral types in Fig.~\ref{fig:SpTDist}, distinguishing between RSGs with and without IR excess. For the sample with IR excess (hereafter, the ``dusty sample''), we considered stars satisfying the first and third criteria presented in Sect.~\ref{Data:Selection}, to be consistent with \cite{Bonanos2023VLT}. Typically, RSG atmospheric models have $\textrm{m}_{3.6}-\textrm{m}_{4.5}$ ranging from $-0.3$ to $-0.15$ from synthetic photometry. The selected limit of an IR excess $\geq 0.1$~mag is therefore conservative. We indicate spectral types of the dusty sample in the right panel of Fig.~\ref{AbsCMD}. Although most dusty RSGs are of late-type (i.e. more evolved), we classified a few very dusty RSGs as K-types (see Sect.~\ref{SecDusty} for a discussion).

\subsection{SED luminosities} \label{SecResLumi}

We collected photometry from \textit{Gaia}~DR3 \citep{GaiaDR3}, SkyMapper DR2 \citep{Skymapper}, Pan-STARRS1 \citep{Chambers2016}, 2MASS \citep{Cutri2003}, VISTA Hemisphere Survey \citep{McMahon2012}, and \textit{Spitzer} IRAC bands \citep{Fazio2004,Rieke2004}. We presented these magnitudes for each source in Table~\ref{PhotoTable}. We used these magnitudes to construct the SED for each RSG. We examined each SED individually and discarded bad photometric points when they were significantly contaminated by excess flux from another source (e.g. due to a low instrumental spatial resolution). For the 97 targets with sufficient photometric data in the optical, near-IR, and mid-IR, we calculated the luminosities using one of three methods: \textit{i)} when the range from 0.4 to 8.0 $\mu$m was densely populated with a plethora of photometric points, we integrated the observed SED, resulting in a fully data-driven stellar luminosity. Secondly, \textit{ii)} when the near-IR magnitudes were missing, we fitted a Planck function to the photometric points and obtained the luminosity by integrating the Planck function. Finally, \textit{iii)} when a $K_S$-band magnitude was available, we used bolometric corrections from \citet[][$BC_{\rm K} = 2.69\pm 0.11$~mag for K-type stars and $BC_{\rm K} = 2.81\pm 0.07$~mag for M-type stars]{Davies2018} to obtain the absolute magnitude and subsequently, the luminosity. 

Given that RSG fluxes peak in the near-IR, integrating the observed SED (first method) underestimates the luminosity in cases where near-IR photometry was not available. Furthermore, the third method was not applicable in most cases, given the lack of $K_S$-band photometry and the increased likelihood of contamination to the $K_S$-band magnitude for more distant sources. Ultimately, we have derived the luminosities of 97 RSGs using the second method. We find that 45 RSGs have a luminosity above log$(L/$L$_{\odot}) = 5.0$. We used the first and third methods on 34 and 73 RSGs, respectively, to verify a general agreement among these three methods within error bars. Examples of SED fits with a Planck function are shown in Fig.~\ref{fig:HRD_Lfits} (see Sects.~\ref{SecHRD} and \ref{SecDusty} for a further discussion on these objects). We corrected the magnitudes for the Galactic foreground extinction from \citet{Schlafly2011}\footnote{We used an online tool for each of the galaxies \url{http://ned.ipac.caltech.edu/byname}}. We did not correct for extinction inside the host galaxy, and therefore the luminosities should be considered as lower limits. We also note that in distant galaxies, despite discarding bad photometry, the mid-IR photometry occasionally still appeared slightly contaminated. The mid-IR fluxes, however, only have a minor contribution to the total luminosity of the RSG. The luminosities are presented in the penultimate column of Table~\ref{ParamsTable}. The uncertainty in the luminosity was obtained by propagating the magnitude and distance uncertainty of each galaxy. The other columns of Table~\ref{ParamsTable} contain the properties derived from spectral fitting and are described in Sect.~\ref{SecResSpec}.

\subsection{Light curves}\label{Data:LCs}
\subsubsection{Optical light curves} \label{SecVarOpt}
\begin{figure*}[h]
\begin{center}
    \centerline{\includegraphics[width=1.6\columnwidth]{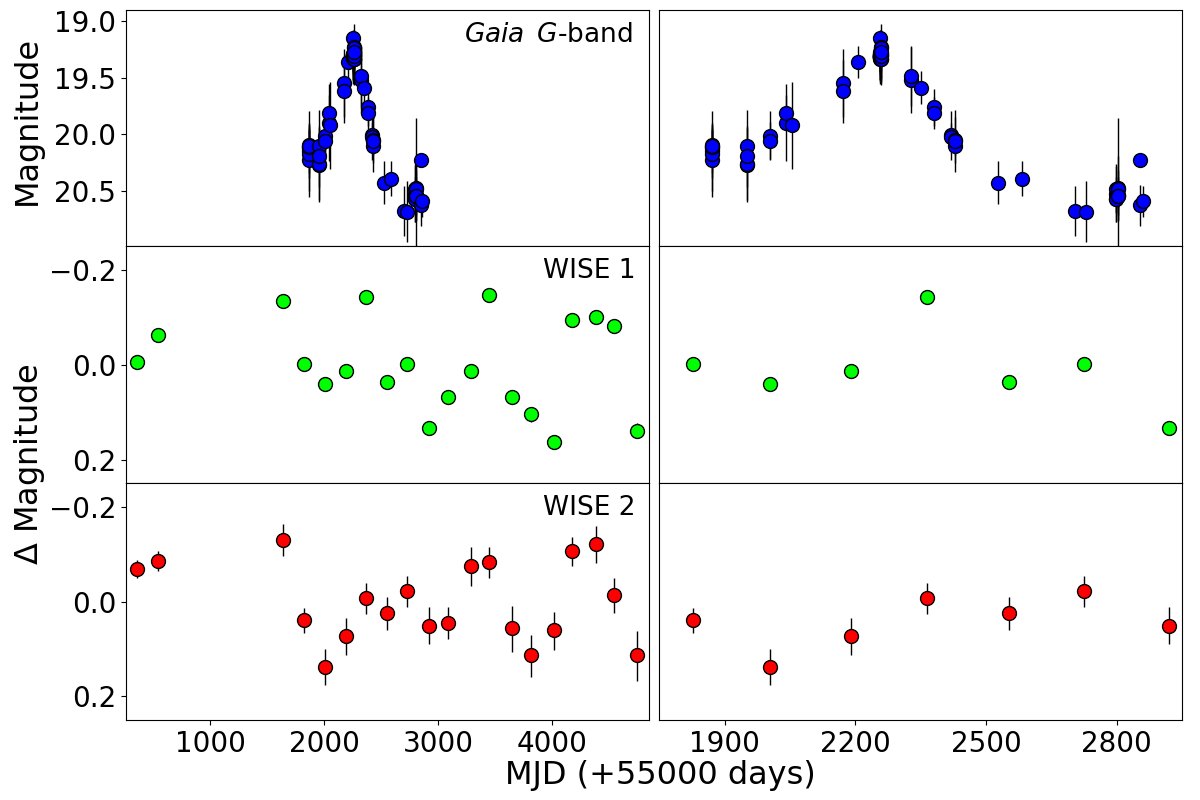}}
    \caption{\textit{Gaia} $G$-band (top), WISE1 (middle) and WISE2 (bottom) light curves for NGC300-59. A zoom-in around the peak of the \textit{Gaia} measurements is shown on the right.}
    \label{fig:NGC300-1LC}
\end{center}
\end{figure*}

RSGs may vary up to two magnitudes in the optical, particularly if they are very evolved and luminous \citep{Kiss2006, Yang2011, Yang2012}. We, therefore, searched the PanSTARRS, \textit{Gaia} and ATLAS \citep{Tonry2018} surveys for multi-epoch photometry of our targets to construct light curves and study them for variability. Due to the faintness of our targets, we were only able to obtain optical light curves for 7 targets from ATLAS, 21 targets from PanSTARRS~DR2, and 4 targets from \textit{Gaia}~DR3. We found ATLAS light curves in both the ATLAS-c and ATLAS-o bands for NGC55-40, NGC55-75, NGC55-135, NGC55-165, NGC300-263, NGC300-154, and NGC3109-167 (see last column of Table~\ref{PhotoTable}), with hundreds of data points covering $\sim$7.5~yrs. However, the error bars on the data were too large to indicate any variability from the light curves. 

From PanSTARRS~DR2, we found multi-epoch photometry for 21 targets in 5 galaxies (the other 4 galaxies were too south to be covered by PanSTARRS~DR2), with at least two measurements in one band with an uncertainty below 0.1~mag (see Fig.~\ref{fig:AllPS1} for all PanSTARRS~DR2 light curves). We computed the amplitude of the light curve by taking the difference between the minimum and maximum values. The last column of Table~\ref{PhotoTable} lists the photometric bands for which these targets have two or more measurements. Only four targets (NGC3109-167, SextansA-1906, NGC247-154, NGC253-872), have coverage in all \textit{grizy} bands. Here, we highlight five interesting cases with an amplitude greater than 1~mag. NGC253-222 has $\Delta i,z,y$ of $\sim$1.8~mag $\sim$1.4~mag and $\sim$2.5~mag respectively, NGC253-872 has $\Delta r$ of $\sim$1.4~mag, NGC253-1534 has $\Delta i$ of $\sim$2.1~mag, NGC247-447 has $\Delta i$ of $\sim$1.0~mag and NGC247-154 has $\Delta r,i$ of $\sim$1.4~mag and $\sim$1.0~mag, respectively. Each of these five RSGs was classified as an M-type RSG, with three of them showing IR excess.

For four targets (NGC300-59, NGC300-125, NGC247-3683, NGC253-872), we constructed light curves from \textit{Gaia}~DR3, containing 70 data points spanning $\sim$~3 years. For these sources, we reported a min-max $\Delta G$ of $\sim$1.5, $\sim$0.9, $\sim$0.6 and $\sim$0.6~mag respectively. We show the \textit{Gaia}~DR3 light curve of NGC300-59 in the top panel of Fig.~\ref{fig:NGC300-1LC} (see Fig.~\ref{fig:AllGaia} for the other 3 light curves). The stellar luminosity for this source was log$(L/$L$_{\odot}) = 5.21$. NGC300-59 was particularly interesting as it additionally varied significantly in the mid-IR (see Sect.~\ref{SecVarIR}). The observed long period of $\sim 1000$~days (min to max $\sim 500$~days) is reasonable compared to the periods found for other bright RSGs \citep{Yang2011, Yang2012}. 

Recently, \cite{Riello2021} developed a method to infer variability of \textit{Gaia} sources for which a light curve is not available. This method hinges on the fact that magnitudes have a typical, expected uncertainty corresponding to their brightness. When a RSG has a higher $G$-band uncertainty than expected for a given brightness, it is likely due to fluctuating measurements (due to e.g. intrinsic variability). For 93 RSGs in our sample containing \textit{Gaia} $G$-band measurements, we tested whether they are above the threshold indicated in Fig.~14 of \cite{Riello2021}. We find 38 RSGs above this threshold, indicating they are `likely' medium-to-high amplitude variables. From the previous eight RSGs highlighted as either PanSTARRS~DR2 or \textit{Gaia} variables, five (NGC300-59, NGC300-125, NGC247-154, NGC253-872 and NGC253-222) had a \textit{Gaia} uncertainty above the threshold and one had no \textit{Gaia} measurements, meaning only two sources were not correctly selected as variables using this method. NGC247-3683 did not exceed the threshold, albeit at the limit, despite showing a 0.5~mag variation in its \textit{Gaia} $G$-band light curve.

\subsubsection{Mid-infrared light curves} \label{SecVarIR}
\begin{figure}[h]
\begin{center}
    \centerline{\includegraphics[width=1.05\columnwidth]{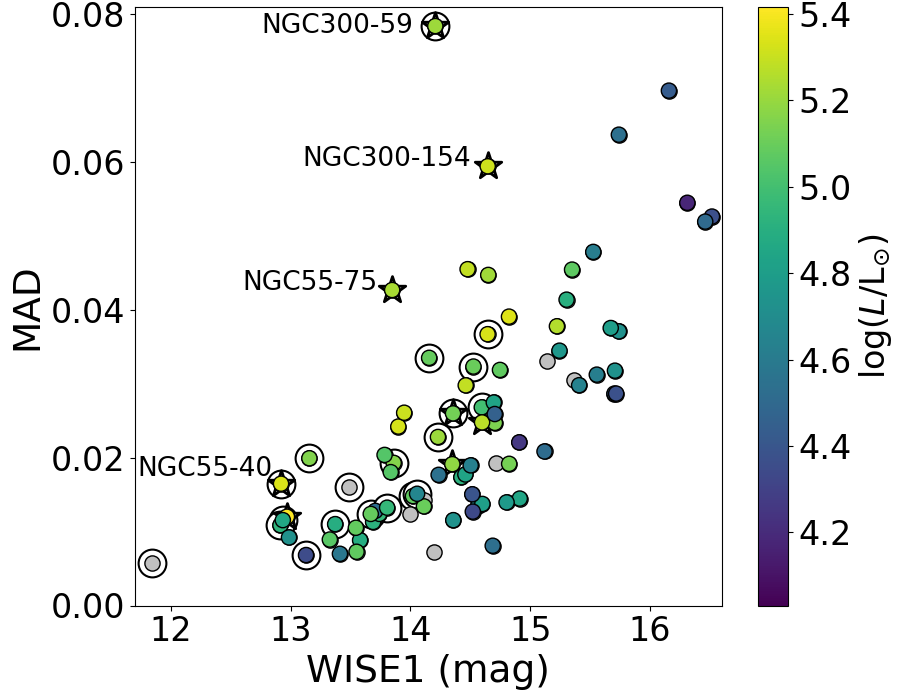}} 
    \caption{WISE1 versus MAD$_{\rm WISE1}$ diagram for RSGs in NGC~55 and NGC~300. The color bar indicates the luminosity for each RSG and the gray dots indicate cases where a luminosity was not obtained. Black stars indicate eight RSGs that were reported as variable in optical surveys. Open circles indicate dusty RSGs. Four RSGs with both optical variability and a high MAD value relative to their WISE1 magnitude are labeled.}
    \label{MADvsL}
\end{center}
\end{figure}

If a RSG loses a variable amount of mass with time, it should become apparent in the mid-IR light curve, as the circumstellar dust, on which the photons scatter, forms and dissipates as it propagates. We have collected light curves for WISE1 (3.4~$\mu$m) and WISE2 (4.6$~\mu$m) for NGC~55 and NGC~300 (86 sources), using the difference imaging pipeline for WISE data described in \citet{De2023}. Where possible, using these light curves, we studied the connection between the optical and mid-IR variability. All WISE1 light curves contained 20 epochs, spanning $\sim$12.5 years. First, we derived the median absolute deviation (MAD) values for WISE1, to check the relation between variability and luminosity. We present these MAD values as MAD$_{\rm W1}$ in Table~\ref{PhotoTable}. In Fig.~\ref{MADvsL} we visualized the sources that are variable in the mid-IR in a MAD diagram, as a function of their magnitude and luminosity. We see that the MAD value generally increased with magnitude, due to the increasing uncertainty on the photometry. For sources with comparable magnitudes, we showed that RSGs with higher luminosities (see the color bar in Fig.~\ref{MADvsL}) generally have a higher MAD value, which is in agreement with past studies \citep[see e.g.,][]{Yang2018, Antoniadis2024}. Similarly, we found that dusty RSGs generally have a higher MAD value than those without dust, populating the upper parts of the curve shown in Fig.~\ref{MADvsL}.

In Fig.~\ref{MADvsL}, we highlighted eight sources that were reported as variable from their optical light curves (see Sect.~\ref{SecVarOpt}). All of these RSGs had luminosities above log$(L/$L$_{\odot}) \sim 5.0$. We note NGC300-59 (M2-M4~I), which showed the highest MAD value from its WISE1 light curve and was additionally reported as a high-amplitude optical variable in \textit{Gaia}~DR3. In Fig.~\ref{fig:NGC300-1LC} we show the correlated WISE1 and WISE2 light curves of this source in the middle and bottom panel, respectively. From the zoom-in (right panels of Fig.~\ref{fig:NGC300-1LC}), we found that the WISE1 light curve peaks $\sim$100 days after the peak of the \textit{Gaia} light curve, although the cadence of the WISE observations prevented us from making a more precise estimate of the delay. We discuss this dusty RSG further in Sect.~\ref{SecDusty}.

\section{Methodology} \label{SecMeth}
\subsection{The grid of \textsc{marcs} models} \label{SecMARC}
We have constructed a grid of 1D \textsc{marcs} model atmospheres \citep{Gustafsson2008} to model for the spectroscopic properties of our RSG sample. Typically, spherical models are used, with a theoretical mass of $M_{\rm RSG}=15\,\textrm{M}_{\odot}$ \citep[see e.g.,][for a discussion on this assumption]{Davies2010}. Additionally, we adopted the default value for the microturbulent velocity, $\xi=5\,\textrm{km s}^{-1}$, as \cite{Gonzalez2021} have shown that the fitted effective temperatures are relatively unaffected by $\xi$ ($\leq 20$~K). Since the targeted galaxies widely varied in metallicity ($Z$), we have adopted a grid of models spanning from log$(Z/\rm Z_{\odot})=+0.25$~to~$-1.00\,\textrm{dex}$ with $\Delta$log$(Z/\rm Z_{\odot})=-0.25\,\textrm{dex}$. We then used linear interpolation, to make the metallicity grid denser to a grid step of $\Delta$log$(Z/\rm Z_{\odot})=-0.125\,\textrm{dex}$. Since the TiO molecular bands display a strong degeneracy between $T_{\rm eff}$ and $Z$, it was not feasible to derive both parameters from optical spectra. Typically, one resorts to $J$ or $K_S$-band spectroscopy to robustly derive a value for $Z$ \citep[e.g.,][]{Gazak2015, Patrick2015, Patrick2017}. Therefore, to lift the degeneracy between $T_{\rm eff}$ and $Z$, we fixed the metallicity to the average metallicity reported in the literature for each galaxy (see Table~\ref{GalaxyTable}). Lastly, we fixed the surface gravity, $\rm log\,\it g$, to +0.0$\,\textrm{dex}$ \citep[commonly assumed in e.g.,][]{Carr2000, Levesque2005, Davies2010}. We emphasize that the resulting best-fit parameters should not be affected by this choice, given that no spectral features in our wavelength range were sensitive to changes in $\rm log\,\it g$ (e.g., the Ca~\textsc{ii} triplet).

We have modeled $T_{\rm eff}$, the color excess E$(B-V)$, and $v_{\rm rad}$ as free parameters. Through linear interpolation, we constructed a $T_{\rm eff}$ grid ranging from $3300-4000\,\textrm{K}$ with $\Delta T_{\rm eff}=10\,\textrm{K}$ and from $4000-4500\,\textrm{K}$ with $\Delta T_{\rm eff}=25\,\textrm{K}$. The changing step size at $T_{\rm eff} \geq 4000\,\textrm{K}$ was motivated by the decreasing impact of the temperature on the strengths of the TiO band as the temperature increases. Although constraining the temperature down to a $10\,\textrm{K}$ precision was not realistic, given all the physical uncertainties on the RSG temperature (e.g., spectral variability due to hysteresis loops), a dense grid allowed to constrain the statistical uncertainties to greater accuracy. To derive E$(B-V)$, and subsequently $A_V$, we applied \citet{Fitzpatrick1999} extinction laws to the models. We assumed a total-to-selective extinction coefficient of $R_V = 3.1$ for all galaxies. Although we note that $R_V$ may be different in a circumstellar environment compared to the interstellar medium \citep{Massey2005}, we argue that fixing $R_V$ to this value was acceptable, as \cite{Gonzalez2021} demonstrated that the resulting $T_{\rm eff}$ is barely affected by changes in $R_V$. We varied E$(B-V)$ between 0.0 and 3.0 magnitudes with a stepsize of $\Delta$E$(B-V) =0.05\,\textrm{mag}$.

\subsection{Spectral fitting strategy} \label{SecFit}
First, we scaled the spectral properties of the \textsc{marcs} models to match those of the observed spectra. We downgraded the resolving power of the models to $R \sim 1000$ (using \textsc{PyAstronomy.instrBroadGaussFast}), to match the spectral resolution of our FORS2 grism. We then resampled the wavelength grid of the model, such that it was identical to the grid of the spectrum. To optimize the modeling of the RSG spectra, a similar smoothing tactic was chosen as presented in \cite{DeWit2023}. We fitted 15 to 20 selected spectral windows of $\sim 50\AA$, containing crucial spectroscopic diagnostics (i.e. the molecular TiO bands to derive $T_{\rm eff}$). By selecting bins, regions without diagnostics did not influence the resulting best-fit $\chi^2$. Moreover, fitting narrow wavelength bins exposed some of the inherent uncertainties in specific rotational and vibrational line transitions of the TiO molecules present in the models. Both of these effects on the best-fit $\chi^2$ were now minimized as they are smoothed out over a $\sim 50\AA$ range. The number of bins used per star depended on the range of the observed spectrum. Some spectra missed a TiO band either in the reddest or bluest part (i.e. either due to reduction errors, due to the physical position of the slit at the edge of the detector, or due to contamination of blue wavelengths by another source). In a few spectra, some TiO bands were substantially contaminated with strong nebular emission lines from the ambient medium, and as such were excluded from the calculation. We calculated $\chi^2$ as follows:

\begin{ceqn}
\begin{align}
    \chi^2 = \sum^n_i \left( \frac{F_i - F_{i,\textrm {mod}}}{E_i}\right)^2,
    \label{eq:chisq}
\end{align}
\end{ceqn}

\noindent in which $n$ is the amount of bins, $F_i$ is the flux of the spectral bin, $F_{i, \textrm {mod}}$ the flux of the model bin and $E_i$ is the uncertainty on the flux in spectral bin $i$. The model with the lowest $\chi^2$ was accepted as the best-fit model, and its free parameters were accepted as the best-fit numerical solution.

We iteratively fitted the entire RSG sample using the previously discussed minimum $\chi^2$ routine. First, we fixed $Z$ and $v_{\rm rad}$ to the value of the host galaxy (see Table~\ref{GalaxyTable} for details per galaxy). We then fitted the 2D grid of \textsc{marcs} models to obtain the best-fit $T_{\rm eff}$ and E$(B-V)$. However, the resulting best-fit parameters marginally depend on the adopted values for $v_{\rm rad}$, especially in galaxies with a strong rotational curve \citep[e.g., NGC~253, see][]{Hlavacek2011}. Therefore, using a cross-correlation technique, we used the best-fit model as an input to derive a more accurate value for $v_{\rm rad}$. Following a similar $\chi^2$ algorithm, we shifted the model with $\Delta v_{\rm rad}=10$~km s$^{-1}$ to obtain the numerical best-fit velocity shift from the lowest $\chi^2$. When the best-fit $v_{\rm rad}$ differed from the adopted $v_{\rm rad}$ of the host galaxy, we updated the input $v_{\rm rad}$ and repeated the fitting process until the solution converged. When converged, we verified the numerical results by visually inspecting the best-fit model applied to the spectrum. Similar to \cite{DeWit2023}, we constructed a 2D $\chi^2$~map to derive the 1~to~3$\sigma$ uncertainties from contours, enclosing the $\chi^2 = \chi^2_{\rm min}+\Delta\chi^2$ values (where $\Delta\chi^2 = 2.3$, 6.2 and 11.8 are the values for which the probabilities exceed 1$\sigma$-3$\sigma$, respectively).

\subsection{The grid of evolutionary models} \label{SecPosydon}
To compare the results from spectroscopic fitting to evolutionary predictions, we constructed single-star populations of RSGs at different metallicities using the state-of-the-art \textsc{Posydon} population synthesis framework \citep{Fragos2023}. \textsc{Posydon} builds on top of \textsc{mesa} models \citep{Paxton2011,Paxton2013,Paxton2015,Paxton2018}. We assume a \cite{Kroupa} initial mass function and a constant star formation history, to generate a sample of RSGs. In our population synthesis analysis, those that produce a neutron star as the end product of stellar evolution ($\sim 7-30$~M$_{\odot}$) are considered RSGs.

Although most massive stars are expected to interact with a binary companion \citep{Sana2012, Mink2014}, in this first comparison we ignore the possible binary history of RSGs, producing populations based solely on single star tracks. For this, we used the currently developed default \textsc{Posydon} tracks extending below solar metallicity (Andrews et al., in prep.), at $Z~=~0.1,~0.2,~0.45,~1.0~Z_{\odot}$. We added one metallicity ($Z = 0.3~Z_{\odot}$) to the existing options, as the metallicity of NGC~55 (42~RSGs) is neither well represented by $Z~=~0.2~Z_{\odot}$ nor $Z~=~0.45~Z_{\odot}$.

Several assumptions in \textsc{Posydon} are tailored specifically to massive stars. For the convective core overshooting, default assumptions for \textsc{Posydon} were assumed. The exponential overshooting scheme was used for a more gradual overshooting, albeit calibrated towards the Bonn step overshoot \citep{Brott2011a}. No overshooting is assumed for shell burning. The models do not consider rotation ($v_{\rm rot}=0~\rm km s^{-1}$). The wind scheme incorporated in these models follows the "Dutch" wind recipes \citep[][for cool and hot winds respectively]{Jager1988, Vink2001}. The mixing length parameter has been fixed to $\alpha_{\rm MLT} = 1.93$. This parameter handles the convection physics in the RSG atmosphere and the envelope structure (and subsequently, the stellar radius and effective temperature) is therefore highly dependent on the assumed value.

To generate a robust enough RSG population, we generate 50k seeds. We can then obtain useful statistics such as the average effective temperature of a population at given metallicity <$T_{\rm eff}$> (later incorporated into Fig.~\ref{fig:teffmethods}) according to the current evolutionary predictions and assumptions.

\begin{figure*}[h!]%
    \centering
    {{\includegraphics[width=2.05\columnwidth]{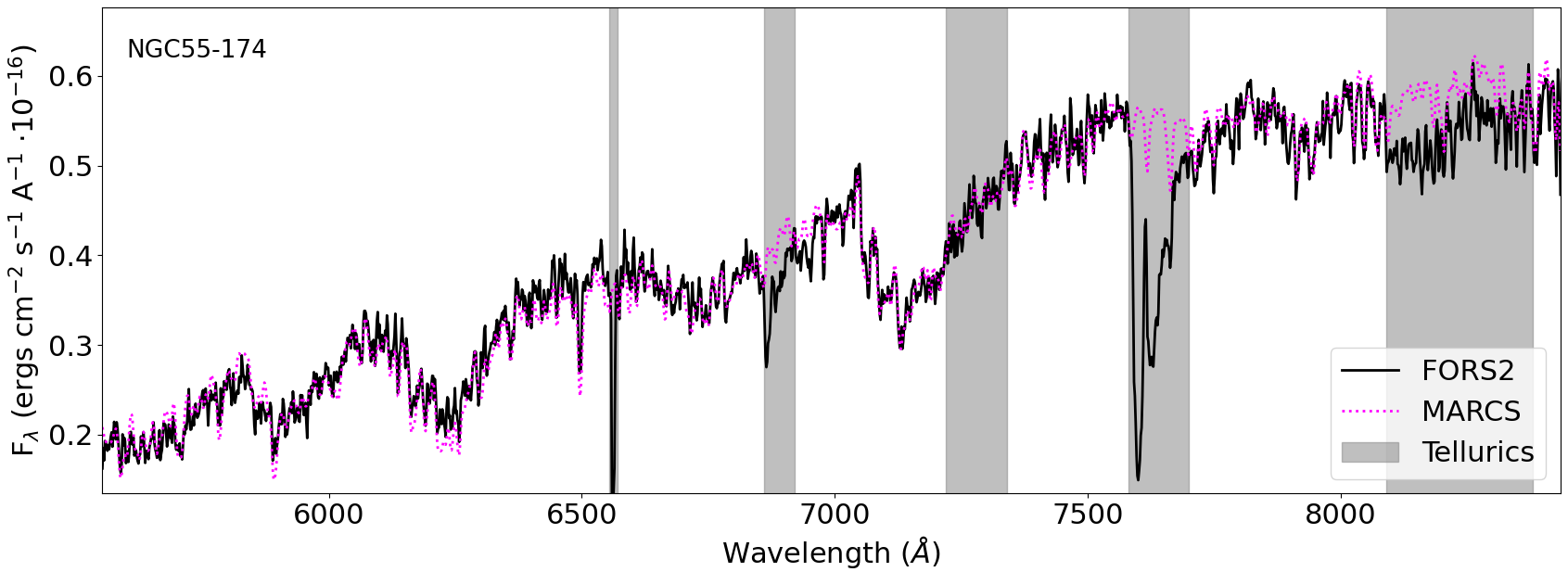} }}%
    \caption{Best-fit \textsc{marcs} model to NGC55-174, a M0-M2~I star. The FORS2 spectrum is indicated in black, the medium-resolution \textsc{marcs} model in magenta and telluric contaminated regions in grey shades.}%
    \label{fig:bestfit}%
\end{figure*}

\begin{figure}[h!]%
    \centering
    {{\includegraphics[width=1\columnwidth]{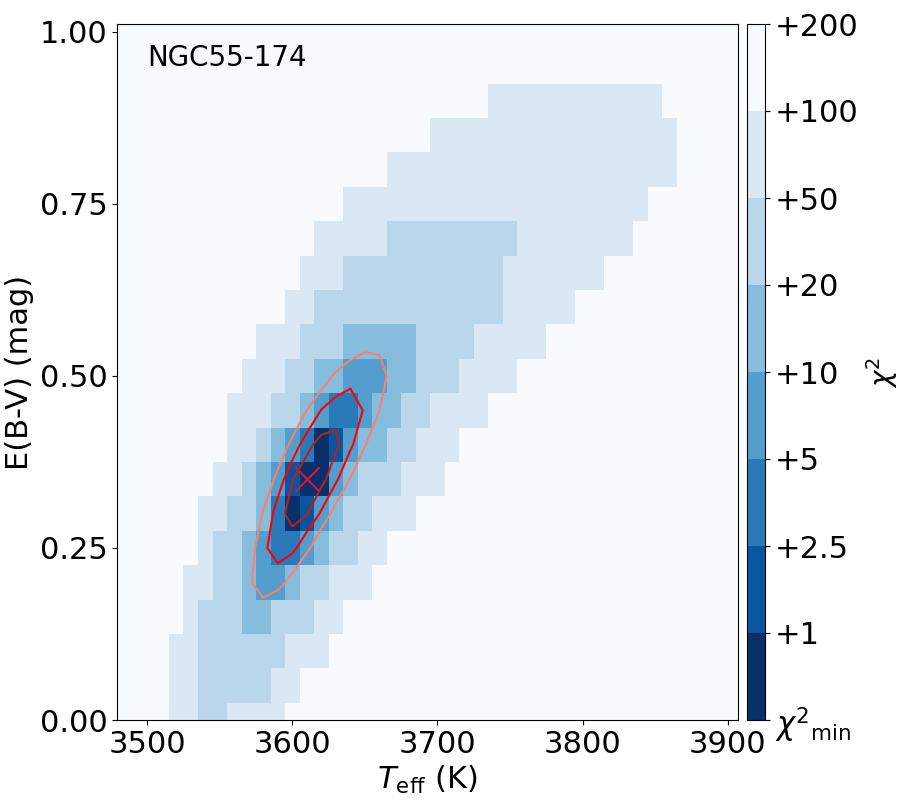} }}%
    \caption{Uncertainties of the best-fit $T_{\rm eff}$ and E$(B-V)$ of NGC55-174. The red cross indicates the best-fit solution. The blue shades indicate a decrease in goodness-of-fit, with the 1$\sigma$-3$\sigma$ limits shown with red contours.}%
     \label{fig:chisqmap}%
\end{figure}

\begin{figure}[h]
\begin{center}
\centering
    \includegraphics[width=1\columnwidth]{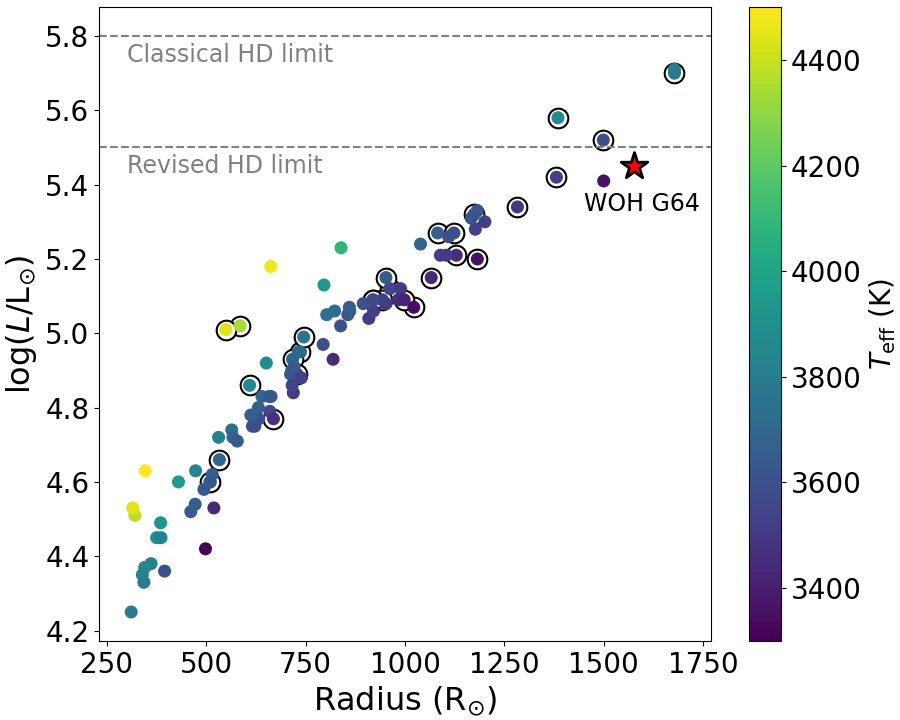}
    \caption{Stellar radius versus luminosity for 97 RSGs. Circles indicate sources from the dusty sample, four of which (including 3 dusty ones) exceed the revised Humphreys-Davidson limit. Note that the most luminous point contains two overlapping extreme RSGs. WOH~G64 is indicated for reference.}
    \label{fig:RadvsL}
\end{center}
\end{figure}

\section{Results} \label{SecRes}
\subsection{Spectroscopic results}
\label{SecResSpec}

\begin{figure}[h!]
\begin{center}
    \centerline{\includegraphics[width=1\columnwidth]{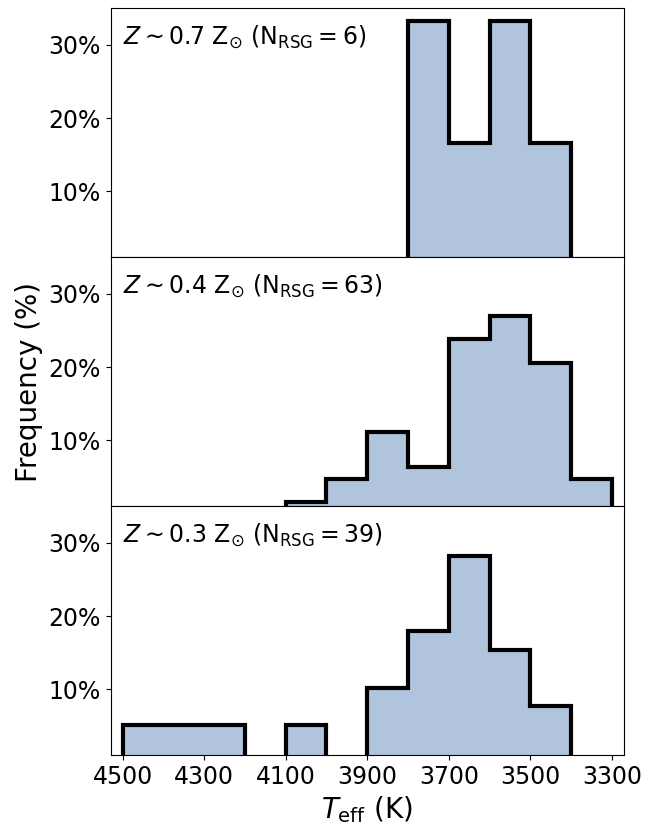}} 
    \caption{Normalized histogram of the $T_{\rm eff}$ distribution of our RSG sample for $0.7~\rm Z_{\odot}$ (NGC~253; top),  $0.4~\rm Z_{\odot}$ (NGC~300, NGC~247 and NGC~7793; middle), and $0.3~\rm Z_{\odot}$ (NGC~55 and NGC~1313; bottom), excluding sources with flags 2-4.}
    \label{fig:hists}
\end{center}
\end{figure}

\begin{figure*}[h]
\begin{center}
    \centerline{\includegraphics[width=1.8\columnwidth]{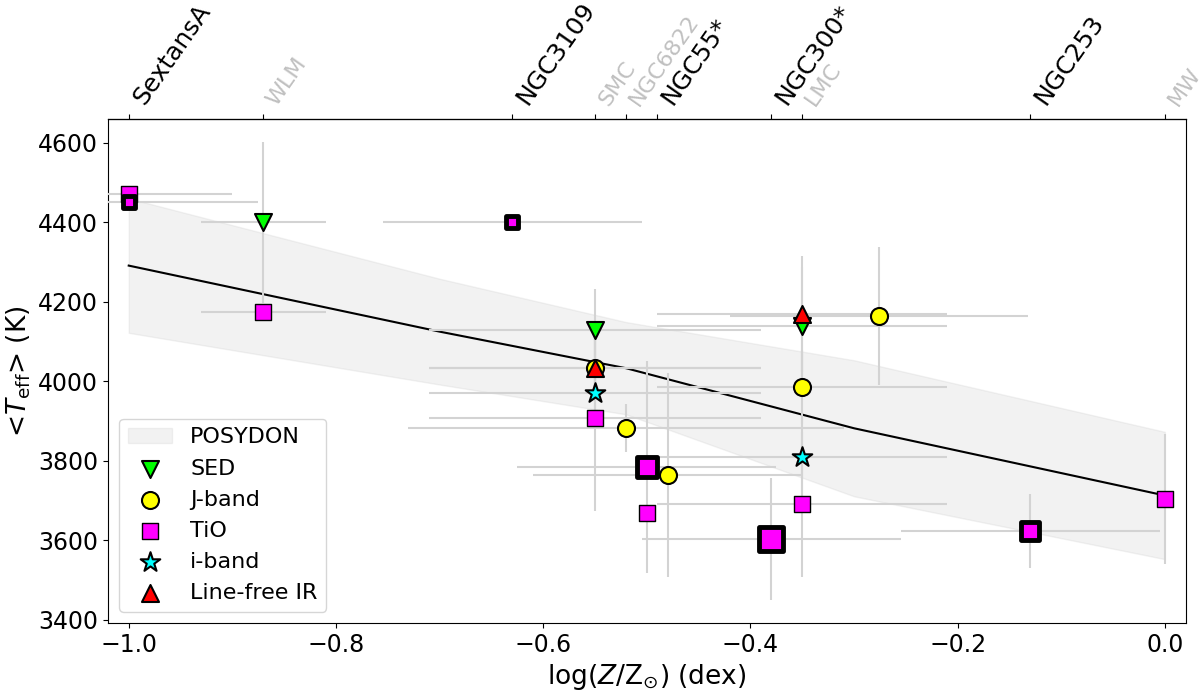}} 
    \caption{Comparison of our modeled $T_{\rm eff}$ with the average $T_{\rm eff}$ of literature studies of RSGs based on line-free (red triangles), SED methods (green inverted triangles), $J$-band spectral lines (yellow circles), TiO bands (magenta squares) the $i$-band spectral lines (cyan stars), at a range of metallicities. TiO results from this work are marked with a thick border; larger squares represent larger sample sizes. The black solid line is the mean $T_{\rm eff}$ of a population of RSGs based on default \textsc{Posydon} single star tracks, with the grey shades indicating up to one standard deviation. NGC~300$^\ast$ includes NGC~300, NGC~247, and NGC~7793; NGC~55$^\ast$ includes NGC~55 and NGC~1313.\\
    }
    \label{fig:teffmethods}
\end{center}
\end{figure*}

We have applied the grid of \textsc{marcs} models to 127~RSG spectra and obtained their physical properties. In most cases we found a good fit (numerically and visually) to these diagnostics: \textit{(i)} the TiO absorption bands for $T_{\rm eff}$, \textit{(ii)} the slope of the spectrum for E$(B-V)$ and \textit{(iii)} individual spectral lines such as Ba~\textsc{ii}~$\lambda$6496, to verify the computed $v_{\rm rad}$. We demonstrate the quality of a single best-fit model applied to NGC55-174 in Fig.~\ref{fig:bestfit} (Fig.~\ref{fig:AllMarcs} presents all the other fitted spectra). The uncertainties of the parameters from the \textsc{marcs} model fits were computed using the $\chi^2$~map presented in Fig.~\ref{fig:chisqmap}, by adopting the 1$\sigma$ contour as the final uncertainty on $T_{\rm eff}$ and E$(B-V)$. 

In Table~\ref{ParamsTable}, we present the ID, quality flags, the best-fit $T_{\rm eff}$, best-fit E$(B-V)$, adopted log$(Z/\rm Z_{\odot})$, and best-fit $v_{\rm rad}$ for 127 RSGs, as well as both the stellar luminosities computed in Sect.~\ref{SecResLumi} and the stellar radii of 97 RSGs using the Stefan-Boltzmann relation. We show the radii and luminosities of each RSG in Fig.~\ref{fig:RadvsL}. We find 6 RSGs with extreme stellar radii ($R \gtrsim 1400 \,\ \rm R_{\odot}$; NGC7793-34, NGC300-125, NGC247-154, NGC247-155, NGC253-222 and NGC1313-310), which are discussed further in Sect.~\ref{SecDusty}. Their SED-fitted luminosities are shown in Fig.~\ref{fig:HRD_Lfits}. Each of these SED fits is good by eye (except NGC247-155), although future observations are necessary to show that these are indeed single, uncontaminated RSGs.

\begin{table*}
\centering     
  \begin{threeparttable}
    \caption{Physical parameters of our RSG sample.}
    \tiny
    \label{ParamsTable}
    \renewcommand{\arraystretch}{1.3}
        \begin{tabular}{l l l l c r c l}        
        \hline                 
        ID & Flag & $T_{\rm eff,TiO}$ & E$(B-V)$ & log($Z/\rm Z_{\odot}$) & $v_{\rm rad}$ & log($L/\rm L_{\odot}$) & $R/\rm R_{\odot}$ \\ 
         & & (K) & (mag) & (dex) & (km s$^{-1}$) & (dex) & (dex) \\    
        \hline                       
        NGC55-40	&		&	3480$^{+0}_{-10}$ & 0.4$^{+0.0}_{-0.1}$ & $-$0.50& 140	&	5.34	$\pm$	0.05 & 1285$^{+85}_{-75}$ \\
        NGC55-75	&		&	4050$^{+175}_{-70}$ & 0.6$^{+0.1}_{-0.05}$ & $-$0.50& 230	&	5.23	$\pm$	0.05 & 840$^{+75}_{-110}$\\
        NGC55-87	&		&	3510$^{+0}_{-10}$ & 0.35$^{+0.0}_{-0.05}$ & $-$0.50& 210	&	5.09	$\pm$	0.07 & 945$^{+85}_{-75}$\\
        NGC55-93	&	2	&	3620$^{+10}_{-10}$ & 0.5$^{+0.05}_{-0.05}$ & $-$0.50& 210 	&	5.15	$\pm$	0.02 & 950$^{+35}_{-30}$\\
        NGC55-135	&	0	&	3540$^{+10}_{-10}$ & 0.25$^{+0.05}_{-0.0}$ & $-$0.50& 80 	&	5.12	$\pm$	0.06 & 960$^{+65}_{-65}$\\
        NGC55-146	&		&	3560$^{+10}_{-10}$ & 0.2$^{+0.05}_{-0.05}$ & $-$0.50& 230 	&	5.09	$\pm$	0.02 & 920$^{+25}_{-25}$\\
        NGC55-147	&		&	3530$^{+100}_{-60}$ & 0.35$^{+0.55}_{-0.35}$ & $-$0.50& 110 	&	-	&	-	\\
        NGC55-149	&		&	3670$^{+80}_{-40}$ & 0.6$^{+0.15}_{-0.1}$ & $-$0.50& 60 	&	4.95	$\pm$	0.02 & 735$^{+40}_{-50}$\\
        NGC55-152	&		&	3590$^{+20}_{-10}$ & 0.35$^{+0.05}_{-0.05}$ & $-$0.50& 90 	&	5.08	$\pm$	0.03 & 895$^{+35}_{-40}$\\
        NGC55-165	&		&	4425$^{+75}_{-25}$ & 0.3$^{+0.05}_{-0.0}$ & $-$0.50& 200 	&	5.18	$\pm$	0.05 & 660$^{+50}_{-60}$\\
        NGC55-174	&		&	3610$^{+20}_{-10}$ & 0.35$^{+0.05}_{-0.05}$ & $-$0.50& 170 	&	5.05	$\pm$	0.04 & 855$^{+40}_{-45}$\\
        NGC55-194	&		&	3690$^{+50}_{-50}$ & 0.25$^{+0.1}_{-0.15}$ & $-$0.50& 130 	&	4.95	$\pm$	0.03 & 730$^{+45}_{-40}$\\
        NGC55-200	&		&	3490$^{+10}_{-10}$ & 0.3$^{+0.1}_{-0.05}$ & $-$0.50& 180 	&	5.04	$\pm$	0.03 & 910$^{+35}_{-35}$\\
        NGC55-202	&		&	3830$^{+110}_{-80}$ & 0.1$^{+0.1}_{-0.1}$ & $-$0.50& 70 	&	4.86	$\pm$	0.05 & 610$^{+60}_{-65}$\\
        NGC55-216	&		&	3730$^{+40}_{-50}$ & 0.3$^{+0.05}_{-0.1}$ & $-$0.50& 150 	&	5.05	$\pm$	0.08 & 805$^{+100}_{-90}$\\
                \vdots	&	\vdots	 & \vdots	&	\vdots	&	\vdots &	\vdots &	\vdots\\
        \hline
        \end{tabular}
    \begin{tablenotes}
        \small
        \vspace{5pt}
        \item Notes: This table is available in its entirety in machine-readable and Virtual Observatory (VO) forms in the online journal. A portion of 15 rows is shown here for guidance regarding its form and content. \\
        Flags: (0) red and blue TiO bands show slightly discrepant temperatures; (1) fitted by eye; (2) fits not satisfactory after visual inspection; (3) blue excess flux (wavelengths below $\lambda \leq 6000\AA$ were discarded); (4) low signal-to-noise ratio skews the uncertainties to extreme values.
    \end{tablenotes}
    \end{threeparttable}
\end{table*}

For five RSG spectra (NGC55-93, NGC300-125, NGC247-447, NGC247-2706, and NGC253-784), the best-fit model selected by the $\chi^2$ routine did not provide a good solution. From these five, we fitted only the spectrum of NGC300-125 by eye, as there were better solutions available. We measure $T_{\rm eff} = 3350$~K and E$(B-V) = 0.4$~mag, with a $v_{\rm rad}$ of 160~km s$^{-1}$, adopt conservative uncertainties of 50~K and flag this source in Table~\ref{ParamsTable} ($\rm flag=1$). We found a good $\chi^2$ in the other four cases, yet, the best-fit model still fitted unsatisfactorily by eye ($\rm flag=2$).

For eight RSG spectra, the flux in the blue wavelengths was contaminated, yielding unreasonable E$(B-V)$ when included in the fit, and subsequently, unreasonable $T_{\rm eff}$. For these cases (NGC7793-274, NGC7793-331, NGC55-1861, NGC247-155, NGC247-1825, NGC253-1534, NGC253-222, NGC253-1226), we discard the blue wavelengths ($\lambda \leq 6000\AA$) and fit only wavelengths where the RSG dominates the flux. The derived extinction for these eight sources ($\rm flag=3$) should be taken cautiously. For four RSG spectra, the fit was satisfactory (NGC55-447, NGC55-NC2, NGC253-3509, and NGC253-1137), but the uncertainties were extreme due to the low signal-to-noise ratio of the spectra ($\rm flag=4$). Finally, for five RSG spectra (NGC55-135, NGC300-59, NGC300-154, NGC300-173, NGC300-240), the TiO bands were deeper (i.e. cooler) in the red and shallower (i.e. hotter) in the blue compared to the best-fit model ($\rm flag=0$). However, the effect on the derived $T_{\rm eff}$ is small, and these were therefore included in the analysis in Sect.~\ref{SecDusty}.

To illustrate how the derived temperatures change with metallicity, we show temperature distributions for different $Z$ in Fig.~\ref{fig:hists}, discarding sources with flags 2-4. The peak and temperature spread of the distribution shifts to hotter temperatures with decreasing metallicity (shown by the increasing number of RSGs at hotter temperatures in the middle and bottom panels of Fig.~\ref{fig:hists}). In the bottom panel, we have eight RSGs with a $T_{\rm eff} \geq 4000$~K, compared to only one RSG in the middle panel. Lower metallicity stars have more compact envelopes due to the decreased opacity, therefore, this shift of average temperature with metallicity is expected \citep[see][]{Elias1985, Levesque2017}. Table~\ref{AverageTable} shows the average $T_{\rm eff}$ and $v_{\rm rad}$ for galaxies with $N_{\rm RSG} \geq 5$, as well as the metallicity and sample size. We verified that the average $v_{\rm rad}$ values were in agreement with those listed in Table~\ref{GalaxyTable}.

\cite{Davies2013} showed that the TiO diagnostic produces RSG temperatures that are generally too cool. We have compiled the average temperature of various spectroscopic literature studies on RSG using a variety of diagnostics and compared them to our average $T_{\rm eff}$ in Fig.~\ref{fig:teffmethods}. We present temperatures obtained from SED fits \citep{Gonzalez2021}, metallic lines in the $J$-band \citep{Patrick2015, Patrick2017, Gazak2015, Davies2015}, TiO bands \citep{Levesque2005, Levesque2006, Britavskiy2019b}, metallic lines in the $i$-band \citep{Tabernero2018} and lastly, line-free continuum fits \citep{Davies2013}. Fig.~\ref{fig:teffmethods} demonstrates that the TiO bands systematically yield lower $T_{\rm eff}$, compared to both the predictions from \textsc{Posydon} population synthesis models (see Sect.~\ref{SecPosydon}) and other literature studies. Note that \citet{Gazak2015} and \citet{Patrick2017} derived a mean $Z$ from their $J$-band spectra in NGC~300 and NGC~55, respectively, which slightly differs from the metallicity indicated in the top axis referring to the mean $Z$ of the galaxy listed in Table~\ref{GalaxyTable}.

\begin{table}
\centering     
  \begin{threeparttable}
    \caption{Average parameters of RSGs.}
    \tiny
    \label{AverageTable}
    \renewcommand{\arraystretch}{1.2}
        \begin{tabular}{l l l l r}        
        \hline\hline                 
        Galaxy  & $<T_{\rm eff}>$ & $<v_{\rm rad}>$ & $Z$ & \#\\ 
         & (K) & (km s$^{-1}$) & ($Z_{\odot}$) & RSG\\    
        \hline\hline                        
            NGC~55  & 3785$\pm$270 & 150$\pm$55 & 0.27 & 38 \\
            NGC~247 & 3565$\pm$135 & 170$\pm$50 & 0.40 & 12 \\
            NGC~253 & 3625$\pm$95 & 200$\pm$140 & 0.72 & 6 \\
            NGC~300 & 3600$\pm$160 & 150$\pm$55 & 0.41 & 44 \\
            NGC~7793 & 3675$\pm$125 & 275$\pm$50 & 0.42 & 7 \\
        \hline
        \end{tabular}
    \begin{tablenotes}
        \small
        \vspace{5pt}
        \item Notes: M83, Sextans~A, NGC~3109, and NGC~1313 were excluded from this Table, as only one RSG was modeled in each of these galaxies. Sources flagged with 2, 3 or 4 in Table~\ref{ParamsTable} were excluded to estimate a clean $<T_{\rm eff}>$ for each galaxy.
    \end{tablenotes}
    \end{threeparttable}
\end{table}

\subsection{New color-temperature relations} \label{SecJK}

\subsubsection{Empirical $T_{\rm eff}(J-K_S)$ relations} \label{SecJKemp}

\begin{figure*}[h]
\begin{subfigure}[t]{0.5\textwidth}
\centering
    \includegraphics[width=1\columnwidth]{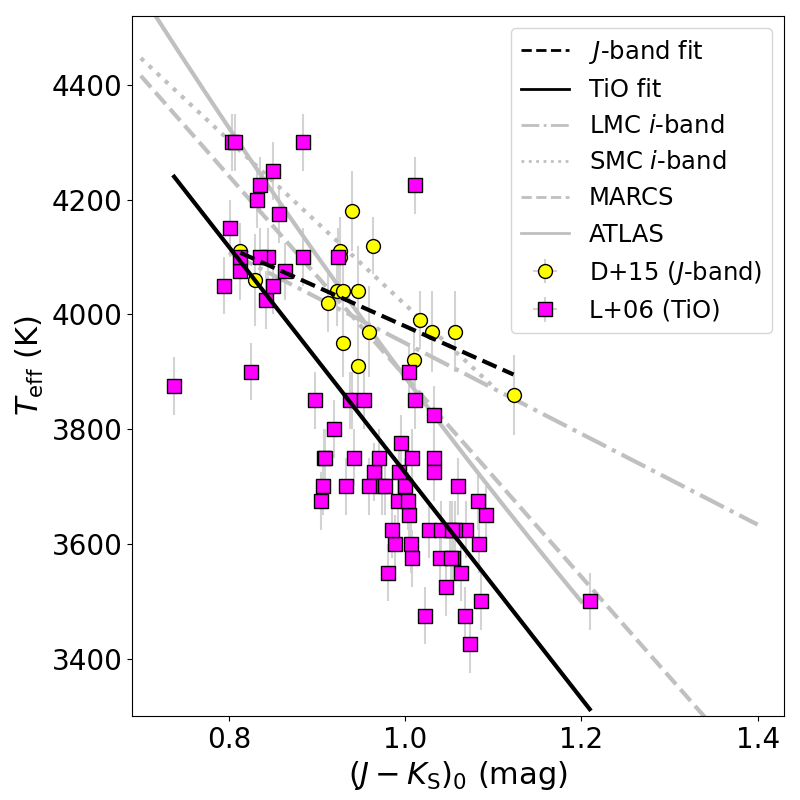}
\end{subfigure}
\begin{subfigure}[t]{0.5\textwidth}
    \includegraphics[width=1\columnwidth]{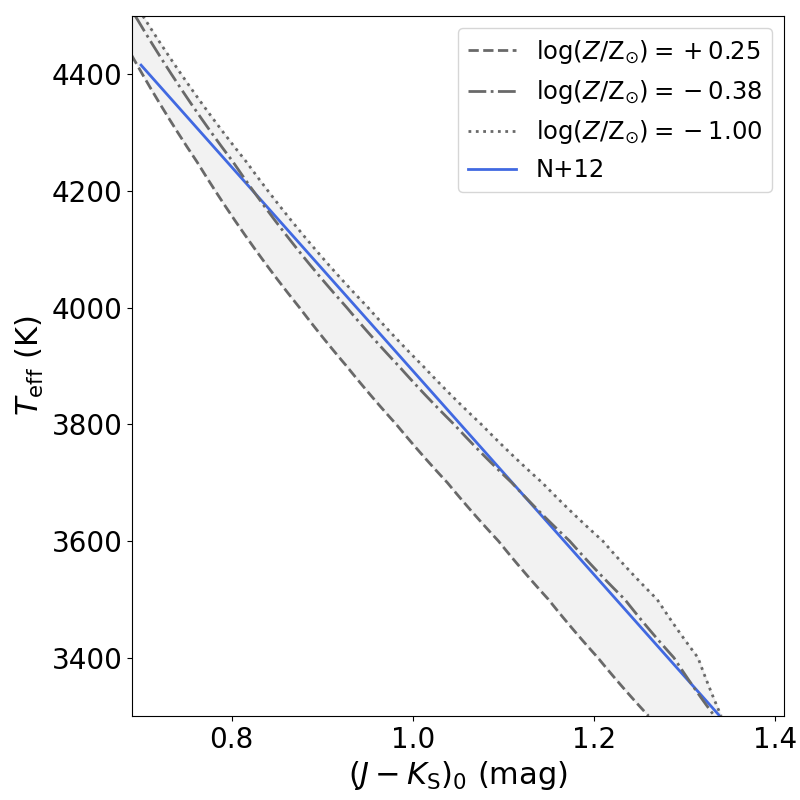}
\end{subfigure}
\caption{\textit{Left}: New empirical $T_{\rm eff}(J-K_S)$ relations compared to  relations from the literature. Black lines present our new empirical relations based on samples from \citet[][squares]{Levesque2006} and \citet[][circles]{Davies2015}. Gray lines indicate empirical relations from \citet[][LMC $i-$band]{Britavskiy2019b} and \citet[][SMC $i-$band]{Britavskiy2019a} and theoretical relations from \citet[][\textsc{marcs}]{Neugent2012} and \citet[][\textsc{atlas}]{Yang2020}. \textit{Right}: The effect of metallicity on the new theoretical $T_{\rm eff}(J-K_S)$ relations. We demonstrate the relation for LMC metallicity with a gray dash-dotted line and the fit from \cite{Neugent2012} with a blue solid line. The gray shaded area indicates the effect of metallicity between the most extreme values of our grid (0.1$-$1.8~$Z_{\odot}$).}
\label{fig:JminK}
\end{figure*}

Color-temperature relations \citep[e.g., $T_{\rm eff}(V-K_S)$,  $T_{\rm eff}(J-K_S)$;][]{Levesque2006, Neugent2012} are often established to estimate the temperature of a population of RSGs from photometry. Such relations require caution, as they are highly influenced by the spectroscopic temperature used for their calibration. Near-IR colors are preferable for such a relation as they are relatively unaffected by extinction and photometric variability, which is much smaller in these bands than in the optical \citep{Schlegel1998, Whitelock2003}. We proceed to derive new $T_{\rm eff}(J-K_S)$ relations calibrated on spectroscopic temperatures from the $J$-band and TiO, similar to those derived by \citet{Britavskiy2019b, Britavskiy2019a} for the LMC and SMC respectively, which were calibrated on the $i$-band.

We combine spectroscopic temperatures of RSGs in the Magellanic Clouds from \cite{Levesque2006} and \citet{Davies2015} to derive new $T_{\rm eff}(J-K_S)$ relations (see Fig.~\ref{fig:JminK}). To derive a relation calibrated on the TiO diagnostics, we used a set of 73 RSG temperatures from \cite{Levesque2006}. For the $J$-band calibrated relation, we used temperatures of 19 RSGs from \cite{Davies2015}. We used the $A_V$ coefficient presented in \citet{Levesque2006} and \citet{Davies2013} to correct the $J-K_S$ color for the effect of extinction. We adopted the following color correction \citet{Schlegel1998}: $(J-K_S)_0 = (J-K_S)- E(J-K) = (J-K_S) - 0.535 E(B-V) = (J-K_S) - 0.535(A_V/3.1)$. The values of $A_V$ are systematically lower in \cite{Levesque2006} compared to those derived in \cite{Davies2013}, due to the use of the TiO diagnostic, and the intrinsic extinction is likely higher. Therefore, the $(J-K_S)_0$ colors for this sample (squares in Fig.~\ref{fig:JminK}) should be interpreted as upper limits, with the true values likely being lower than derived. 

We present the best-fit relations to both sets of RSG temperatures and $(J-K_S)$ colors in both the left panel of Fig.~\ref{fig:JminK} and Table~\ref{RelationsTable}. We compare the derived relations to existing empirical relations \citep{Britavskiy2019b, Britavskiy2019a} and theoretical relations \citep{Yang2020, Neugent2012}. The empirical relations are calibrated by using temperatures derived from $i$-band spectra \citep{Tabernero2018}. The theoretical relations use synthetic photometry from either \textsc{atlas} or \textsc{marcs} models to derive an extinction-free $T_{\rm eff}(J-K_S)_0$ relation. We show the literature relations, the diagnostic on which they are calibrated, the metallicity, and the sample size in Table~\ref{RelationsTable}. The derived $J$-band relation agrees with the $i$-band relation only for LMC metallicity (see Fig.~\ref{fig:JminK}), while the TiO relation agrees much more with the theoretical relations, which predict much steeper dependencies of $T_{\rm eff}$ on $(J-K_S)$. We also note that the SMC \textit{i}-band relation is steeper than its LMC counterpart, suggesting a possible metallicity effect on the slope of the color-temperature relation.

\subsubsection{Theoretical $T_{\rm eff}(J-K_S)$ relations} \label{SecJKsynth}

To investigate whether the changing slope of the color-temperature relations is expected from theoretical models, we derived new relations using the \textsc{marcs} atmospheric models for a range of metallicities (0.1$-$1.8~$Z_{\odot}$). Following the approach of \citet{Neugent2012}, we derived a $(J-K_S)_0$ color for each temperature of the model grid, but for a range of metallicities. We then fit a linear relation to the computed synthetic colors. We find that the slope is relatively unaffected by metallicity (see Fig.~\ref{fig:JminK}, right panel), and that there is a small shift toward bluer colors with increasing metallicity. By fitting the differences in offset and slope for each metallicity on the grid, we determine the dependence of the theoretical $T_{\rm eff}(J-K_S)$ relation on log$(Z/\rm Z_{\odot})$. We find a quadratic dependence of the offset ($a$) and slope ($b$) and present the relations in Table~\ref{RelationsTable}. This new $Z$-dependent relation can be applied to a wider range of metallicities compared to the \cite{Neugent2012} relation. For LMC metallicity, our relation simplifies to the \cite{Neugent2012} relation.

\begin{table*}
\centering     
  \begin{threeparttable}
    \caption{Overview of $T_{\rm eff}(J-K_S)$ relations.}
    \tiny
    \label{RelationsTable}
    \renewcommand{\arraystretch}{1.2}
        \begin{tabular}{l l r l l l}        
        \hline\hline                 
        $T_{\rm eff}$ relation & Calibration method & N$_{\rm sample}$ & log($Z/\rm Z_{\odot}$) & Calibration sample & Reference \\ 
        \hline\hline                        
        $T_{\rm eff} = 4741 -791  (J-K_S)_0$ & $i$-band spectral lines & 188 & $-$0.7 to $-$0.1~dex & \cite{Tabernero2018} & \cite{Britavskiy2019b} \\ 
        $T_{\rm eff} = 5449 -1432  (J-K_S)_0 $ & $i$-band spectral lines & 257 & $-$1.0 to $-$0.4~dex & \cite{Tabernero2018} & \cite{Britavskiy2019a} \\ 
        $T_{\rm eff} = 4664 -684  (J-K_S)_0$ & $J$-band spectral lines & 19 & $-$0.7 to $-$0.1~dex & \cite{Davies2015} & This work\\ 
        $T_{\rm eff} = 5691 -1967  (J-K_S)_0$ & TiO molecular bands & 73 & $-$0.7 to $-$0.3~dex &  \cite{Levesque2006} & This work \\ 
        \hline\hline
        $\log(T_{\rm eff}) =3.82 -0.23  (J-K_S)_0 $ & Synthetic ATLAS & N/A & $-0.85$~dex & \cite{Kurucz2005} & \cite{Yang2020} \\ 
        $T_{\rm eff} = 5638 -1746  (J-K_S)_0$ &  Synthetic MARCS & N/A & $-0.35$~dex & \cite{Gustafsson2008} & \cite{Neugent2012} \\
        $T_{\rm eff} = a^{\ast}+b^{\dagger}  (J-K_S)_0$ &  Synthetic MARCS & N/A & $-$1.0 to $+$0.25~dex & \cite{Gustafsson2008} & This work \\
        \hline
        \end{tabular}
    \begin{tablenotes}
        \small
         \item $^{\ast}a = 5686 + 104 $ log($Z/\rm Z_{\odot}) + 74 $ log($Z/\rm Z_{\odot})^2$, $^{\dagger}b = -1851 - 295 $ log($Z/\rm Z_{\odot})  -166 $ log($Z/\rm Z_{\odot})^2$
        \vspace{5pt}
    \end{tablenotes}
    \end{threeparttable}
\end{table*}

\subsection{TiO scaling relations} \label{SecScaling}

Given that the TiO bands produce RSG temperatures that are too cool, it is evident that a direct comparison between the TiO temperatures derived in Sect.~\ref{SecResSpec} and for example, evolutionary tracks of RSGs cannot be made. To remedy this, we derive linear scaling relations to shift the TiO temperatures to hotter temperatures and perform such a comparison. We calibrate the TiO temperature scaling relation using a set of reliable near-IR temperatures to minimize systematics. We use 19 RSGs in the SMC and LMC from \cite{Davies2015} and \cite{Tabernero2018} to derive the scaling relation, as each of these objects has a $T_{\rm eff,TiO}$, $T_{\rm eff,J}$ and $T_{\rm eff,i}$ measurement. This data set excludes potential differences in spectral reduction or variations in the RSG atmosphere \citep[e.g., hysteresis loops;][]{Kravchenko2019, Kravchenko2021}, as each of the three temperatures were obtained from the same spectrum and the differences in $T_{\rm eff}$ arise solely from the different diagnostics. We argue that the $i$ and $J$-band diagnostics are reliable to benchmark a scaling relation, as they rely on the modeling of metal lines forming near the stellar continuum region. The line profiles in the \textsc{marcs} models are well established near the continuum, with major caveats arising mainly at higher altitudes \citep{Davies2013}. Furthermore, deviations from local thermodynamic equilibrium \citep{Bergemann2012, Bergemann2013, Bergemann2015} have been included in the results of \citet{Davies2015} and \citet{Tabernero2018}.

As the data points from \cite{Davies2015} have asymmetric uncertainties, these need to be properly taken into account when performing the linear fit, to not falsely skew the linear relation. Asymmetric uncertainties are specifically strong for high TiO temperatures, due to the non-linear dependence of the TiO bands on the temperature. The TiO bands significantly increase in strength towards lower atmospheric temperatures, making the upper uncertainty always larger. The most extreme case in our sample has a $\sim$5:1 error ratio. 

Each data point has an uncertainty for both variables, which demands complicated regression techniques \citep[e.g., orthogonal distance regression or ODR;][]{Boggs1990}. However, modern implementations of ODR, such as \texttt{scipy.odr}, only treat symmetric uncertainties, while our case demands the treatment of uncorrelated asymmetric uncertainties \citep{Possolo2019}. As a first step to assess our problem, we generate a cloud of points for each of the 19 data points and their uncertainties. 

The statistically weighted cloud of points will represent the asymmetric uncertainties. The samples for each point are drawn using the Monte Carlo method. We draw from a statistical distribution, where 1$\sigma$ represents the uncertainty on the data point reported in the respective studies. For symmetric errors, this distribution simplifies to a Gaussian distribution. However, for the asymmetric cases, we fitted for the shape of the distribution. We obtained a best-fit skewed distribution using a minimization routine, in which the max likelihood of the best-fit posterior represents the value of the data point, and the 1-$\sigma$ intervals of the best-fit posterior are the asymmetric uncertainties. We then draw 1000 times from the best-fit distribution to generate the final cloud of points. 

\begin{figure*}[h!]
\begin{subfigure}[t]{0.5\textwidth}
\centering
    \includegraphics[width=1\columnwidth]{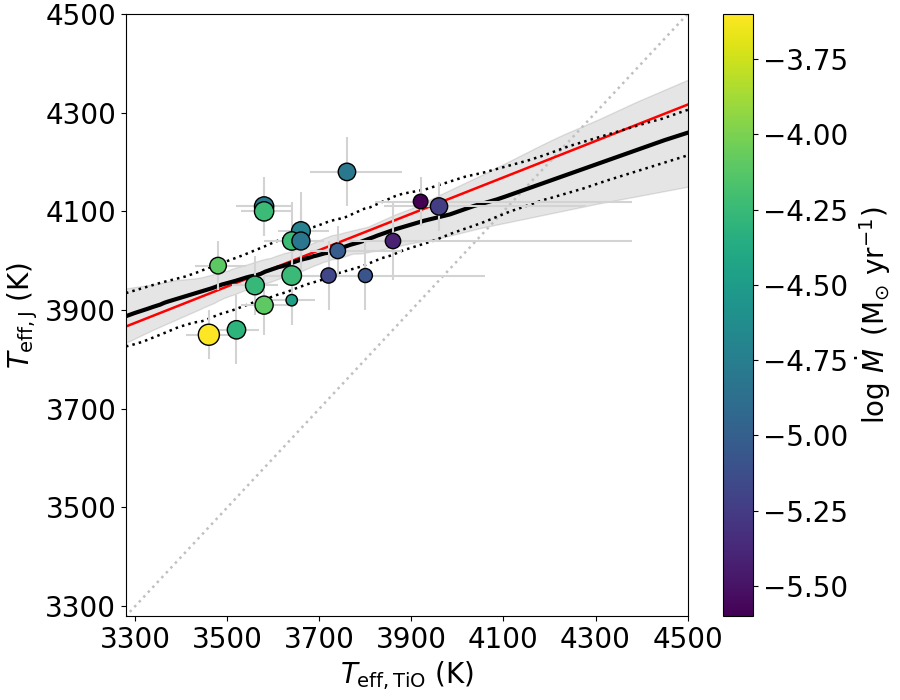}
\end{subfigure}
\begin{subfigure}[t]{0.5\textwidth}
    \includegraphics[width=1\columnwidth]{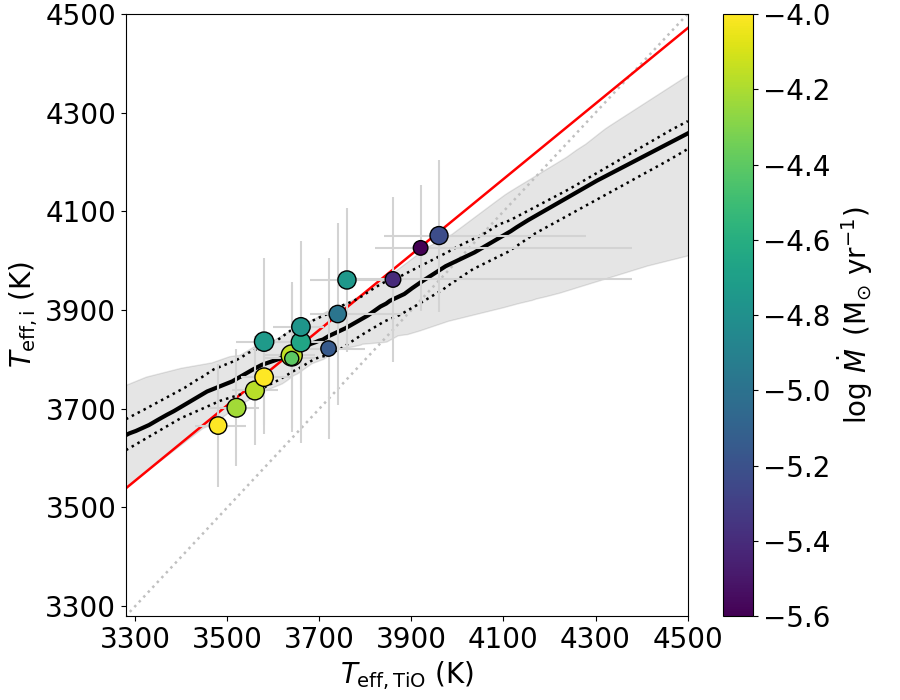}
\end{subfigure}
\caption{Scaling relations between $T_{\rm eff,TiO}$ and $T_{\rm eff,J}$ (\textit{left}) and between $T_{\rm eff,TiO}$ and $T_{\rm eff,i}$ (\textit{right}). The scattered data points and their uncertainties are taken from \cite{Davies2013, Davies2015}. The thick black line indicates the best-fit scaling relation, the gray shaded band shows the combination of slopes and offsets acceptable within a 1$\sigma$ uncertainty and the black dotted lines indicate the scatter on the best-fit relation. In red, we show a simple linear fit to the data when one excludes the effect of asymmetric uncertainties. We show a 1:1 relation for comparison. The color bar indicates the mass-loss rates obtained with \textsc{DUSTY}. The sizes of the markers increase proportionally to metallicity.}
\label{fig:scaling_rels}
\end{figure*}

We use \textsc{ultranest} \citep{Ultranest} to perform a linear fit to the clouds of points (19$\times$1000) to find the best underlying relation. \textsc{Ultranest} is a Bayesian workflow package that uses nested sampling \citep{Skilling2004} to obtain a relation between physical properties. We assume a flat prior and a Gaussian likelihood function, where the mean of the function given by our two-parameter model ($\alpha,\beta$) and a function dispersion $\sigma$. This $\sigma$ is an additional free parameter that describes the scatter along the initial linear model. The resulting scaling relation is parameterized as follows:
\begin{ceqn}
\begin{align}
    T_{\rm eff,new} = \alpha + \beta \times T_{\rm eff,TiO}
    \label{eq:scaling_J}
\end{align}
\end{ceqn}
where for the $J$-band:
\begin{ceqn}
\begin{align*}
    \alpha = 3110^{+400}_{-610},\,\ \beta = 0.24^{+0.17}_{-0.11},\,\ \sigma = 58^{+21}_{-29}
\end{align*}
\end{ceqn}
and for the $i$-band: 

\begin{ceqn}
\begin{align*}
    \alpha = 2020^{+820}_{-750},\,\ \beta = 0.49^{+0.21}_{-0.18},\,\ \sigma = 15^{+31}_{-14}
\end{align*}
\end{ceqn}

\noindent where $\alpha$ indicates the offset, $\beta$ indicates the slope of the relation and $\sigma$ indicates the intrinsic Gaussian scatter. We present the best-fit linear relation to the 19 RSG temperatures in Fig.~\ref{fig:scaling_rels}. In both panels, we show the calibration temperature on the y-axis. The range for which this relation must be applied is dictated by the $T_{\rm eff,TiO}$ points from the literature ($\sim 3450-3950$~K). The shallow slope of the scaling relation suggests that a significant spread in temperatures is not expected for RSGs. Properly sampling the asymmetric uncertainties flattens the best-fit linear relation to the data points compared to doing a simple linear fit (shown in Fig.~\ref{fig:scaling_rels}). The deviation of the temperatures from the 1:1 relation implies that, either the diagnostics are not well represented by the 1D atmospheric models, or they are influenced by other stellar properties (e.g., mass loss). The color bar in Fig.~\ref{fig:scaling_rels} reveals that the temperature scales are indeed less discrepant for lower mass-loss rates. 

\begin{figure}[h!]
\begin{center}
\centering
    \includegraphics[width=1\columnwidth]{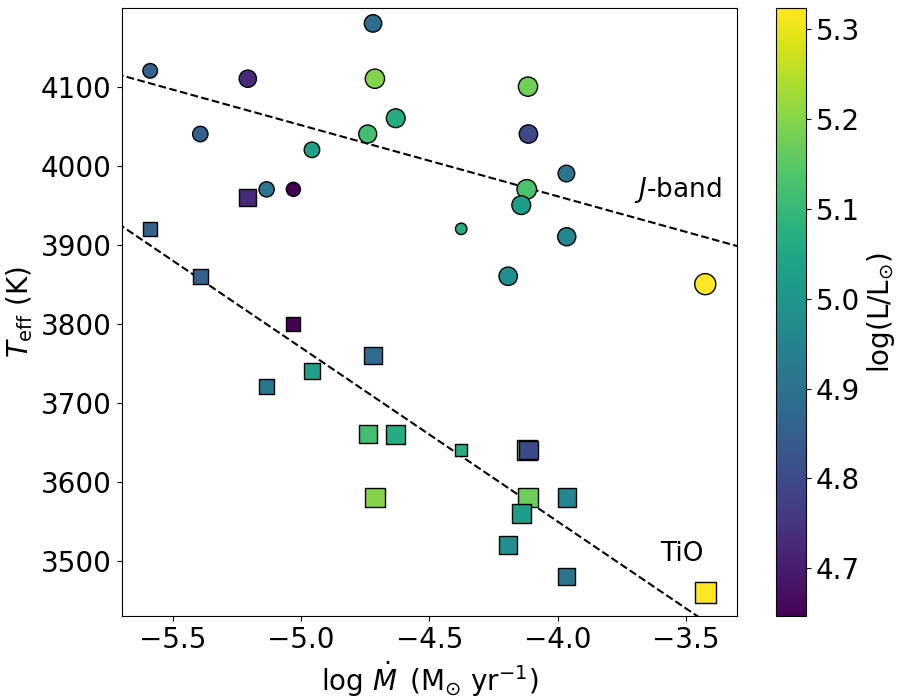}
    \caption{$T_{\rm eff}$ versus mass-loss rate for both TiO temperatures \citep[squares;][]{Davies2013} and $J$-band temperatures \citep[circles;][]{Davies2015}. Color indicates the stellar luminosity; symbol sizes scale with metallicity, as in Fig.~\ref{fig:scaling_rels}. A simple linear fit (dashed line) is shown for guidance.}
    \label{fig:TeffMdot}
\end{center}
\end{figure}

We demonstrate the trend with mass-loss rates in Fig.~\ref{fig:TeffMdot}. The mass-loss rates shown for the SMC sources are taken from \cite{Yang2023}. For the LMC sources, we obtained mass-loss rates by modeling the circumstellar dust shell around the RSG using the radiative transfer code \textsc{DUSTY} \citep{Dusty1}. We assume a spherically symmetric dust shell, which extends to $10^4$ times the inner radius, and a radiatively-driven wind (RDW) as the driving mechanism for mass loss. \citet{Antoniadis2024} discussed caveats in the RDW approach, which when assumed, lead to overestimated mass-loss rates. However, given that the \cite{Yang2023} results are fixed, we assumed RDW winds for the LMC sources regardless, to allow for a direct comparison between the LMC and SMC supergiants. We adopted the same grain size distribution, grid of models, and fitting methodology as defined in \cite{Yang2023}, but adopting a gas-to-dust ratio appropriate for the LMC \citep[$r_\mathrm{gd}=380$;][]{Roman-Duval_2014} and using \textsc{marcs} models with $[Z]=-0.38$. In Fig.~\ref{fig:TeffMdot}, the $J$-band scale is relatively stable with increasing mass-loss rates and shows a large scatter, suggesting that the correlation between the two properties is weak. In contrast, the TiO band scale is more affected and shows a tighter dependence on the mass-loss rate, suggesting they are likely correlated. Using \texttt{scipy.pearsonr}, we find a Pearson coefficient of $= -0.89$ with a $p$-value of $2.5\cdot10^{-7}$, which suggests a tight statistical anti-correlation between the two variables. This observation agrees with the models shown in \cite{Davies2021}, where a strong stellar wind significantly increases the depths of the TiO bands, while spectral lines in the $J$-band remain unaffected. 

For metallicity, we observe a similar effect. Lower metallicity RSGs show a smaller discrepancy between the temperature diagnostics. These RSGs are intrinsically hotter, decreasing the TiO opacity and naturally resolving the discrepancy. Our results strongly suggest that the scaling relation in Eq.~\ref{eq:scaling_J} should be dependent on both metallicity and mass-loss rate. Future data would not only tighten the uncertainty on the linear relation but would also allow for establishing a $Z$ and $\dot M$ dependence. A reliable, all-inclusive scaling relation is particularly useful given that a consistent grid of 3D models or a grid of 1D models, which could self-consistently fit for $\dot M$ has not been established yet \citep[although, see][for pioneering sets of models as well as recent developments regarding these models]{Chiavassa2011, Chiavassa2022, Davies2021, Ahmad2023}.

\section{Discussion} \label{SecDisc}
\subsection{HRD analysis} \label{SecHRD}

\begin{figure}[h]
\begin{center}
\centering
    \includegraphics[width=1\columnwidth]{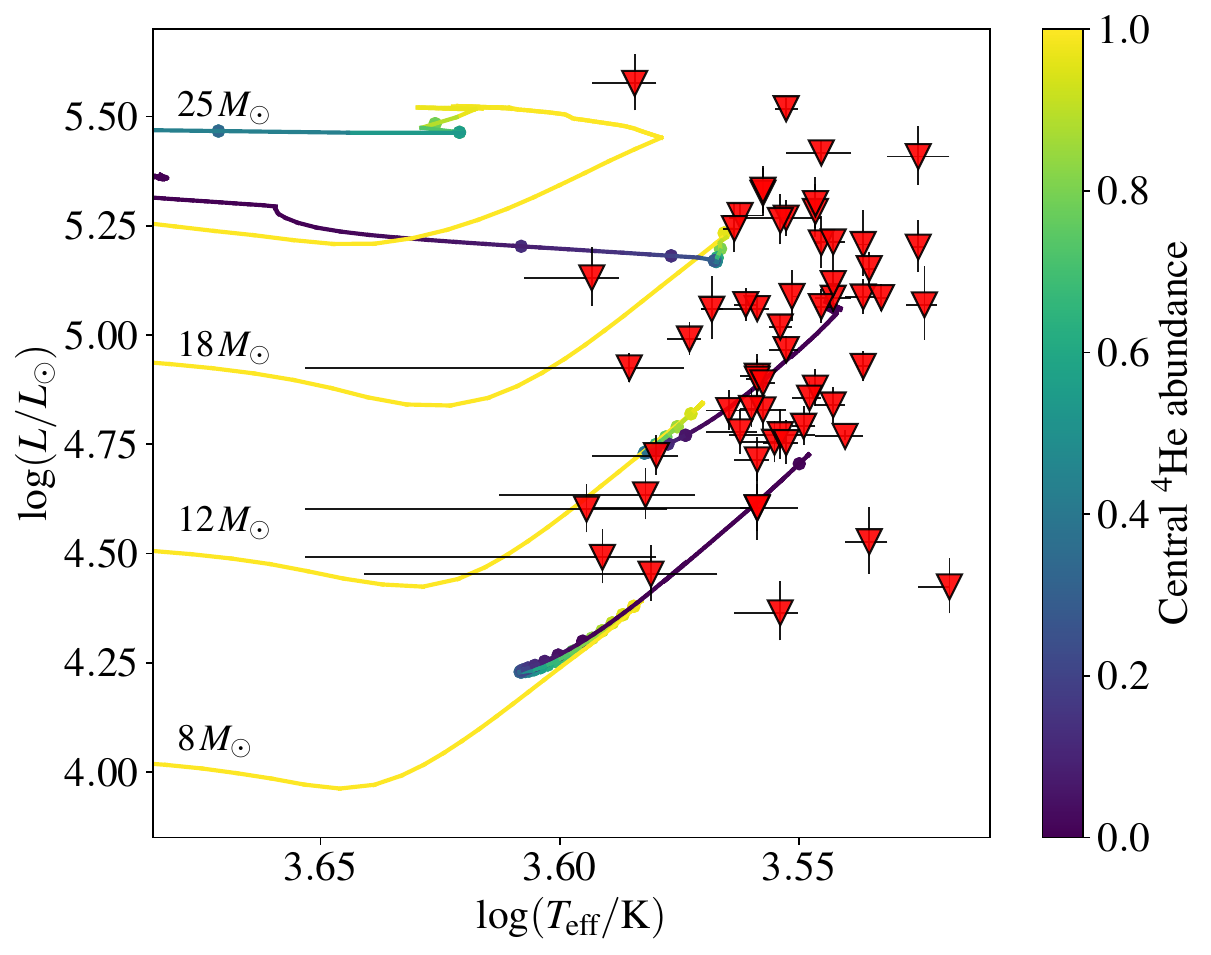}
    \caption{HRD with the observed RSGs (red triangles) in NGC 300, NGC 247, and NGC 7793. Four \textsc{Posydon} tracks from 8$-$25~M$_{\odot}$ computed for log$(Z/\rm Z_{\odot}) = -0.35$ are shown. The color indicates the central helium abundance; the knots on the tracks indicate timesteps of 10,000~yrs.}
    \label{fig:HRD_TiO}
\end{center}
\end{figure}

We construct HRDs and compare our sample to the predictions from \textsc{Posydon} evolutionary tracks computed at log$(Z/\rm Z_{\odot}) = -0.35$ ($0.45~Z_{\odot}$, see Sect.~\ref{SecPosydon}), to match the metallicities of RSGs in NGC~300, NGC~247 and NGC~7793 (log$(Z/\rm Z_{\odot}) = -0.38$). For this metallicity, we obtained both $T_{\rm eff,TiO}$ and log$(L/$L$_{\odot})$ for 59 RSGs. We compare these RSGs to the \textsc{Posydon} tracks in Fig.~\ref{fig:HRD_TiO}. As expected, the data points are not well reproduced by the tracks, and most RSGs are either located at the final stages of the evolutionary track where the RSG evolves quickly towards central iron exhaustion and subsequent core collapse, or occupy the forbidden zone. Two sources, NGC247-154 at log$(L/$L$_{\odot}) \sim 5.52$ and NGC7793-34 at log$(L/$L$_{\odot}) \sim 5.58$ (see Fig.~\ref{fig:HRD_Lfits} for their SED fits), are more luminous than the upper luminosity limit predicted by the evolutionary tracks, even for the core-overshooting properties assumed in \textsc{Posydon}. Although still well below the classical Humphreys-Davidson limit \citep[log$(L/$L$_{\odot}) \sim 5.8$;][]{Humphreys1979}, these two sources have extreme stellar luminosities that exceed the revised Humphreys-Davidson limit \citep[log$(L/$L$_{\odot})\sim 5.5$;][]{Davies2018}. Their SED-integrated luminosities appear satisfactory, however, an inaccurate distance estimate, reddening estimate, or blending and crowding effects may affect the derived luminosity.

\begin{figure*}[h]
\begin{subfigure}[t]{0.5\textwidth}
\centering
    \includegraphics[width=1\columnwidth]{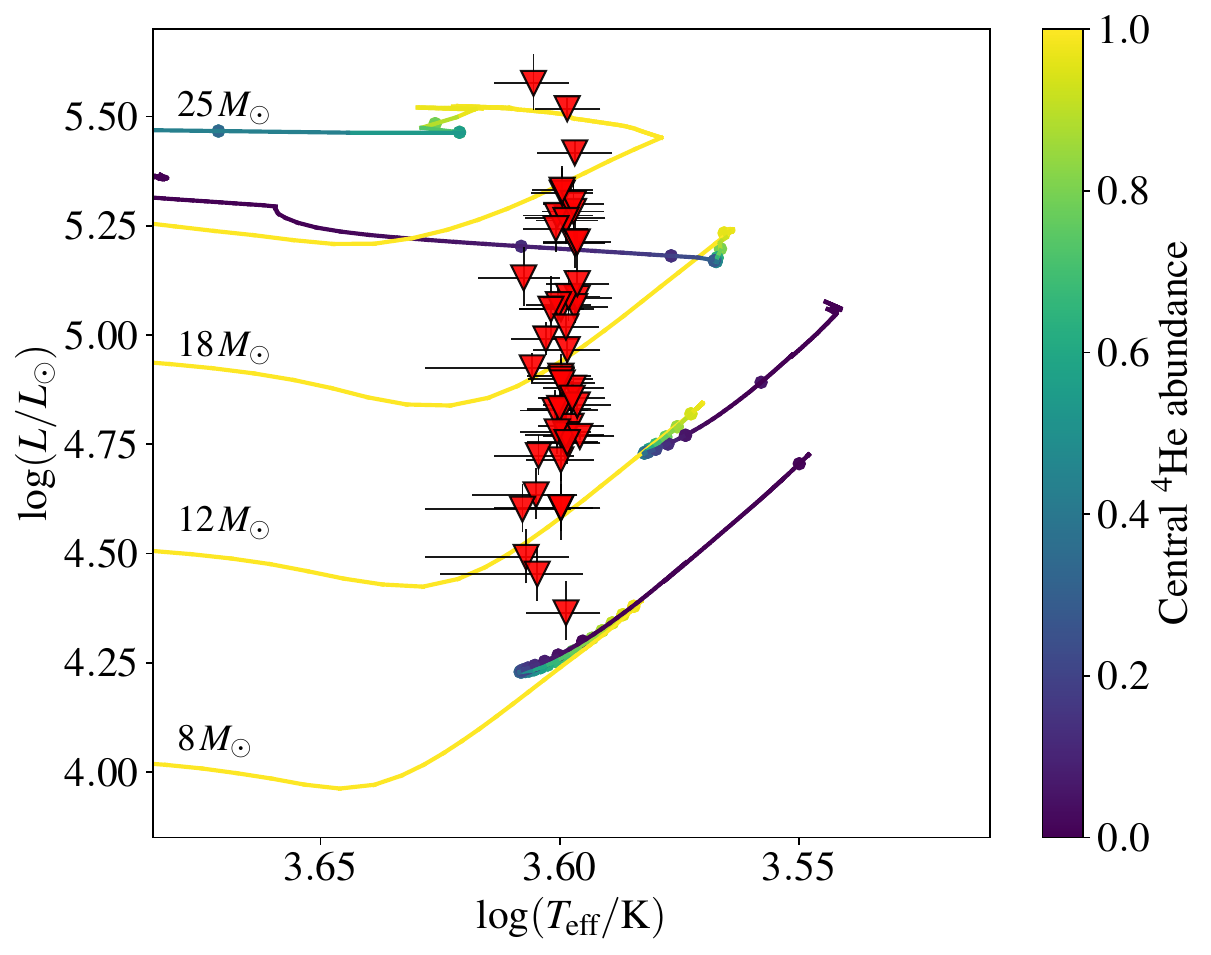}
\end{subfigure}
\begin{subfigure}[t]{0.5\textwidth}
    \includegraphics[width=1\columnwidth]{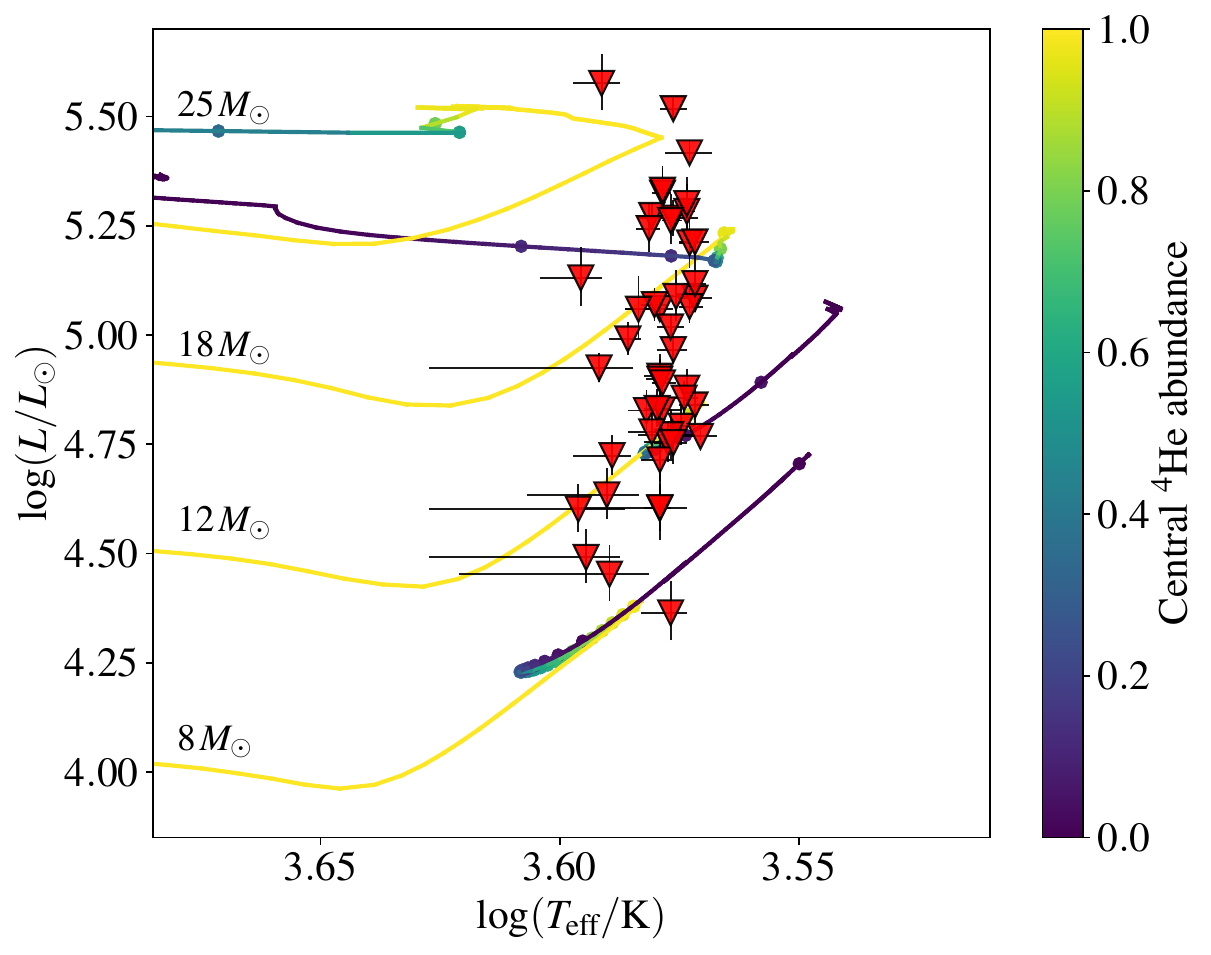}
\end{subfigure}

\caption{Same as Fig.~\ref{fig:HRD_TiO}, but for the observed TiO temperatures scaled by the $J$-band (\textit{left}) and the $i$-band (\textit{right}) relation.}
\label{fig:HRD_rels}
\end{figure*}

We proceed to apply the new scaling relations (see Sect.~\ref{SecScaling}) to 49 out of 59 RSGs shown in Fig.~\ref{fig:HRD_TiO}, which have temperatures within the limits of the scaling relations (3450$-$3950~K), and test which of the relations shift the observed RSGs to where the tracks are in a steady core-helium burning phase. The scaled temperatures, using both scaling relations for these RSGs, are compared to the tracks in Fig.~\ref{fig:HRD_rels}. The $J$-band relation scales the RSG temperatures up to values that are too hot compared to the predicted core helium-burning phase of the Hayashi tracks, except at low luminosities. The RSGs are not expected to populate this part of the HRD, as the \textsc{Posydon} tracks predict a fast Hertzsprung gap crossing for these temperatures. Furthermore, the temperature spread of the $J$-band scaled temperatures is narrower than predicted by the models. The $i$-band scaled temperatures, however, provide an excellent match with the \textsc{Posydon} core helium-burning phase, showing a scatter in log$(T_{\rm eff})$ compatible with the spread of \textsc{Posydon} temperatures. The scaling relations can be further improved by incorporating the effects of mass loss and metallicity. A larger sample of RSG spectra from an instrument such as X-shooter, can yield simultaneous measurements of $T_{\rm eff,TiO}$, $T_{\rm eff,i}$ and $T_{\rm eff,J}$, and tighten the scaling relations further. 

\subsection{Dusty RSGs} \label{SecDusty}
\begin{table}
\centering     
  \begin{threeparttable}
    \caption{Parameters of dusty vs. non-dusty RSGs.}
    \tiny
    \label{Prio_nonPrio}
    \renewcommand{\arraystretch}{1.3}
        \begin{tabular}{l l l l}        
        \hline                 
        Parameter & Dusty & Non-dusty & Diff. \\ 
        \hline                       
        $T_{\rm eff}$ (K) & 3570 (27) & 3630 (84) & $-$60 \\
        $A_V$ (mag) & 1.08 (27) & 0.46 (84) & 0.62 \\
        log($L/\rm L_{\odot}$) (dex) & 5.09 (29) & 4.84 (68) & 0.25 \\
        \hline
        \end{tabular}
    \begin{tablenotes}
        \small
        \vspace{5pt}
        \item Notes: Median values are given; sample sizes are in parenthesis. Sources flagged as 2, 3 or 4 in Table~\ref{ParamsTable} are discarded from the population.
    \end{tablenotes}
    \end{threeparttable}
\end{table}

To better understand the nature of dusty RSGs ($\textrm{m}_{3.6}-\textrm{m}_{4.5}>0.1$~mag and M$_{3.6}\le-9.0$~mag, defined in Sect.~\ref{Data:Class}), we compute their properties and compare them to non-dusty RSGs. Generally, a RSG loses more mass through a stellar wind when it ascends the Hayashi track \citep[e.g., $L$-dependent;][]{Jager1988, Loon2005, Yang2023, Antoniadis2024}. We therefore expect a dusty RSG to be more evolved (i.e. cooler and more luminous). Here, we compute the median effective temperature, extinction, and luminosity of both samples and summarize the results in Table~\ref{Prio_nonPrio}. We assume $R_V=3.1$, to convert the E$(B-V)$ from Table~\ref{ParamsTable} to $A_V$. This assumption is usually applicable for dust in the interstellar medium, but may not hold for circumstellar environments and H~\textsc{ii} regions \citep[e.g., 30 Dor;][]{MaizJesus2014, Brands2023}. For circumstellar material of a RSG, \citet{Massey2005} suggest an $R_V~\geq~4$, which would increase our average $A_V$. As expected, the RSGs in the dusty sample are generally cooler, more luminous, and reach an average optical extinction of around $A_V \sim 1$~mag due to dust scattering. The values presented in Table~\ref{Prio_nonPrio} are therefore explained as a natural result of evolution. The dusty sample also shows a higher degree of variability in the mid-IR (see the encircled sources in Fig.~\ref{MADvsL}), indicating that the circumstellar dust content varies as a result of variable mass loss.

Four RSGs have luminosities exceeding the revised Humphreys-Davidson limit (see Fig.~\ref{fig:RadvsL}). Three of these sources are dusty (NGC7793-34, NGC247-154, and NGC253-222), with NGC1313-310 showing a small IR excess ($\textrm{m}_{3.6}-\textrm{m}_{4.5}\sim0.0$~mag). Lastly, the dusty RSG NGC247-155 is just below the revised luminosity limit and has a stellar radius of $R \sim 1400 \,\ \rm R_{\odot}$. However, both the luminosity and the temperature estimates are flagged for this star. These sources are potential distant analogs of WOH~G64, although tighter constraints on the luminosity are needed to verify their extreme nature.

Almost half of the classified RSGs (16 out of 33, or 48$\%$) in the dusty sample were assigned the latest spectral type (M2-M4~I). This is another indication that a RSG is very evolved \citep{Davies2013} and, therefore, more likely to have a high mass-loss rate, enriching the circumstellar environment with dust. Indeed, 15 out of these 16 RSGs have log$(L/$L$_{\odot}) \geq 5.0$. In contrast with the dusty sample, the fraction of M2-M4~I stars in the non-dusty sample is only 27$\%$. The non-dusty distribution peaks at spectral type M0-M2~I, with 46$\%$ of the RSGs assigned this type. The fraction of classified K-types is higher for the non-dusty sample (28$\%$) than for the dusty sample (12$\%$). 

NGC300-59 and NGC55-40 are members of the dusty M2-M4~I sample and show variability in both the optical and mid-IR (see Fig.~\ref{MADvsL}). The strong variability across multiple wavelengths, their late spectral type, and their luminosity (log$(L/$L$_{\odot}) = 5.21$ and log$(L/$L$_{\odot}) = 5.34$, respectively) all suggest that these are very massive and evolved RSG. These stars are reminiscent of the progenitor of SN 2023ixf in terms of luminosity and mid-IR variability \citep{Soraisam2023, Kilpatrick2023, Jencson2023}. Furthermore, three out of five RSGs that show at least one magnitude of variability in PanSTARRS~DR2 (NGC253-222, NGC247-154, and NGC247-447), are dusty. All of these have log$(L/$L$_{\odot}) > 5.2$. Optical light curves for the remaining 11 dusty RSGs do not currently exist. However, we expect multi-epoch photometry to be available for our RSG sample in the next decade, from upcoming variability surveys, such as the \textit{Rubin} Observatory Legacy Survey of Space and Time.

RSGs from the dusty sample potentially lose so much mass, that they may become self-obscured due to dust formation. A definition of a dust-enshrouded RSG was recently given by \citet{Beasor2022}, for sources with both $A_V > 2$~mag and an additional corresponding excess at $\lambda \geq 2\mu$m. In their study, they find two dust enshrouded RSGs ($A_V = 2.48$ and $A_V = 7.92$~mag), both with $\tau_V \geq 7$. For these sources, this implies that less than $\sim 0.1 \%$ of the $V$-band photons penetrate the dust shell, whereas a less strict definition of $\sim1\%$ of the photons emerging from the shell, would correspond to $\tau_V \sim 4.6$. Therefore, we suggest that criteria based on the degree of obscuration should be considered for defining dust-enshrouded RSGs. In our sample, we have two candidate dust-enshrouded RSGs satisfying the criteria of \citet{Beasor2022}, NGC55-244 with $A_V=2.17$~mag and NGC300-173 with $A_V=2.33$~mag. We find a candidate dust-enshrouded RSG fraction of 1.6$\%$, whereas \citet{Beasor2022} report a fraction of around 3$\%$. However, as argued in \citet{Davies2013}, constraining the extinction from optical \textsc{marcs} model fitting causes $A_V$ values to be underestimated. Therefore, we consider our findings to be lower limits.

Five of our sources are flagged as $\rm flag=0$, indicating that the red and blue TiO bands independently produce inconsistent temperatures. Similar inconsistencies have been observed in extreme RSGs, such as WOH~G64, which have been attributed to the presence of circumstellar dust \citep{Levesque2009}. We have also observed such behavior in [W60]~B90 \citep[log$(L/\rm L_{\odot})=5.45\pm 0.07$;][]{DeWit2023} in the LMC, which seems to be an analog of Betelgeuse, having undergone dimming events similar to the Great Dimming (this RSG is further scrutinized in Munoz-Sanchez et al., in prep.). The Great Dimming was explained by a cool patch on the photosphere followed by dust formation \citep{Levesque2020, Montarges2021, Wheeler2023}, which may also explain the behavior observed in our five sources and [W60]~B90. Three of the $\rm flag=0$ RSGs are dusty (NGC55-135, NGC300-59, NGC300-173), which is in agreement with the observations and conclusions from \citet{Levesque2009}. The other two do not show significant IR excess ($\textrm{m}_{3.6}-\textrm{m}_{4.5}\sim0.0$~mag), but are redder than typical RSG atmospheres expected from synthetic \textsc{marcs} model colors (see the end of Sect.~\ref{Data:Class}).
 
In Fig.~\ref{AbsCMD} we highlighted four dusty K-type RSGs. The spectral classification of M83-682 was complicated by its poor data quality and therefore may not be reliable. The other three K-type RSGs have reliable spectral classifications and IR colors, as they are among the nearest RSGs in the sample. However, the IR excess could be caused by a nearby H~\textsc{II} region. Indeed, the spectrum of NGC55-244 contains a narrow H$\alpha$ emission line, indicating the presence of a contaminating nebula. NGC3109-167 and NGC55-202 do not reveal any emission lines in their spectra and their \textsc{marcs} model fits are satisfactory. Their MAD values are 0.045 and 0.012, respectively, indicating a strong mid-IR variability for NGC3109-167. We found log$(L/$L$_{\odot}) = 5.01$ and log$(L/$L$_{\odot}) = 4.86$ for these sources, respectively, which implies that their mass-loss rates are likely not extreme \citep[e.g.,][]{Jager1988, Loon2005, Goldman2017, Beasor2020, Yang2023} and they are therefore not expected to populate this part of the \textit{Spitzer} CMD. We argue that these two sources are candidate Levesque-Massey variables \citep[see e.g.,][]{Massey2007, Levesque2007, Levesque2009, Levesque2020}. When the \textit{Spitzer} data was taken, these two RSGs could have been in a cooler, more expanded state \citep[such as e.g., WOH~G64;][]{Levesque2009}, where they lose mass at higher rates. This causes a brief epoch in which the IR colors are reddened due to dust formation before the RSGs return to a hotter, more compact state, which is currently observed with spectroscopy.

For NGC55-202, which does not show strong photometric variability, we alternatively propose that it experienced a mass ejection. At the time the \textit{Spitzer} data was taken, it may have partly obscured itself while remaining in a relatively hot state, similarly to the dimming of Betelgeuse. An anisotropic mass ejection outside the line of sight \citep[see e.g.,][]{Kervella2018} would leave the surface unobscured, yet IR excess is expected.
Furthermore, binary mergers for which the primary star is crossing the Hertzsprung-gap \citep[expected to happen in 25$\%$ of the Type II supernova progenitors;][]{Zapartas2019}, which are thought to be responsible for luminous red novae \citep{Rau2007}, may eject envelope material into the circumstellar environment and leave traces of hot dust. In the case of AT2018bwo, the yellow supergiant merger progenitor appeared as a dusty RSG after 1.5 months \citep{Blagorodnova2021}, which could potentially also explain why NGC55-202 appears as a dusty K-type RSG.

\subsection{Color-temperature relations}

The different slopes in the empirically established relations by \citet[][see Sect.~\ref{SecJK}]{Britavskiy2019b, Britavskiy2019a} can not be explained by metallicity alone. A potential explanation could be the effect of a strong stellar wind on the spectral appearance of a RSG. \cite{Davies2021} have shown that a significant stellar wind affects the flux in the $K_S$-band, due to the increasing CO emission emerging from a circumstellar molecular shell. The effects of stellar winds are not incorporated in the \textsc{marcs} models and are therefore not reflected in our theoretical $T_{\rm eff}(J-K_S)$ relation. CO emission starts around $\sim 2.3\mu$m, which is at the edge of the $K_S$-band filter. Here, the efficiency of the filter has already significantly dropped, and the increasing flux due to molecular emission should not affect the $K_S$-band magnitude drastically. We therefore argue that the stellar wind can not explain the change in the slope between the SMC and LMC relations. Another possible explanation may be the improper treatment of 3D processes (e.g. convection) in the 1D atmospheric models, either underpredicting or overpredicting the near-IR fluxes. However, for the example shown in \cite{Davies2013}, fluxes in both $J$ and $K_S$ bands are relatively unaffected by the changes from 1D to 3D models \citep[CO$^5$BOLD;][]{Freytag2002, Chiavassa2011}. We, therefore, argue that the 3D-to-1D simplifications in the \textsc{marcs} models cannot explain the observations either. More investigation is needed to explain the changing $T_{\rm eff}(J-K_S)$ relations with $Z$.

\subsection{TiO bands as a potential $\dot M$ diagnostic} \label{SecTiOMdot}

Throughout this study \citep[and others e.g.,][]{Davies2013, Davies2021}, we have argued that the TiO diagnostic is unsuitable for determining the effective temperature of a RSG. Here, we discuss the TiO bands as a possible diagnostic to probe mass loss. A strong stellar wind impacts the spectral appearance of a RSG spectrum in the optical \citep[drastically so, in cases where $\dot M > 10^{-5} M_{\odot}\mathrm{yr}^{-1}$;][]{Davies2021}. Therefore, if the other parameters that affect the TiO band strengths are known (i.e. $T_{\rm eff}$ and $Z$), they can be fixed in atmospheric modeling, and one can obtain the mass-loss rate. The first step is therefore to constrain $T_{\rm eff}$ and $Z$ independently of the optical spectrum. Near-IR spectra, for example with X-shooter or $K$-band Multi-Object Spectrograph (KMOS), provide spectral lines to constrain both of these properties independently. One can then directly study the dependence of the mass-loss rate on $Z$, which is currently still debated. Furthermore, microturbulent velocities ($v_{\rm turb}$ or $\xi$) can be obtained from the broadening of spectral lines by fitting, for example, \textsc{marcs} model atmospheres \citep[see e.g.,][for an application on $J$-band spectral lines]{Davies2010}. Then, one can study the dependency of $\dot M$ on turbulence. \cite{Kee2021} assumed turbulence to be the dominant mass-loss mechanism driving the RSG stellar wind. Quantifying the relation between turbulence and mass-loss rates using the TiO bands could therefore improve our understanding of the mass-loss mechanism for RSGs. 

The next step would be to extend the \textsc{marcs} models to allow for fitting $\dot M$. Pioneering work by \cite{Davies2021} demonstrated the effect of adding RSG mass loss on top of the hydrostatic \textsc{marcs} model atmospheres, reproducing most features arising from the circumstellar molecular sphere of RSGs \citep[see][for an application to CO transitions in the $K$-band]{GonzalezTora2023b, GonzalezTora2023a}. A grid (for every combination of $T_{\rm eff}$, $Z$, etc.) of such models incorporating stellar wind physics would potentially address several open questions about RSG mass loss. With improved models and tighter constraints on the stellar parameters of RSGs, we argue that the TiO bands can be a valuable mass-loss diagnostic, particularly due to their observational accessibility for RSGs out to a few Mpcs. Many multi-object spectroscopic instruments exist, which could simultaneously observe the TiO bands for large populations of extragalactic RSGs (e.g., FORS2). Low-resolution spectra are sufficient to model the TiO bands, keeping exposure times to a minimum. In contrast, large distances render some existing mass-loss methods unfeasible, such as the CO transitions observed with ALMA \citep[e.g.,][]{Decin2023}. With the excellent spatial resolution and sensitivity of the \textit{James Webb Space Telescope}, SED-fitting of dust emission becomes possible for distant RSGs. However, such modeling still suffers from uncertain, yet high-impact assumptions, such as the gas-to-dust ratio, grain size distribution, and dust composition. We, therefore, conclude that the TiO bands have great potential as a direct $\dot M$ diagnostic, especially at large distances.

\section{Summary and conclusions} \label{SecSum}
We present the physical parameters of the largest sample of spectroscopically studied RSGs (127) outside the Local Group \citep[NGC~55, NGC~247, NGC~253, NGC~300, NGC~1313, NGC3109, NGC~7793, Sextans~A and M83, presented in][]{Bonanos2023VLT}. Using SED-fitting, we derived luminosities for 97 RSGs and found that $\sim 50 \%$ has log$(L/$L$_{\odot}) \geq 5.0$, from which 6 RSGs had extreme stellar radii ($R \gtrsim 1400 \,\ \rm R_{\odot}$). We analyzed optical and mid-IR light curves to characterize RSG variability and found two very evolved RSGs (NGC300-59 and NGC55-40), showing significant variability in both their optical and mid-IR light curves. For all 127 RSG spectra, we presented a refined spectral classification compared to \cite{Bonanos2023VLT}, and performed \textsc{marcs} atmospheric modeling to derive their effective temperatures, extinction factors, and radial velocities.

Temperatures of RSGs are easily obtained through calibrated color-temperature relations, yet existing relations are either applicable to only one metallicity or were calibrated using only one temperature scale. We derived $J-K_s$ color versus temperature relations from \textit{i)} the TiO bands, \textit{ii)} the $J$-band and \textit{iii)} synthetic photometry from the \textsc{marcs} models for a range of metallicities, finding a $Z$-dependent color-temperature relation. A strong $Z$-dependence of the empirical $T_{\rm eff}(J-K_S)$ relations was established by \cite{Britavskiy2019b, Britavskiy2019a}, but is not supported by the theoretical, synthetic relations derived in this work. This hints at an extra contribution to the $J-K_S$ color, which is missed by the physics incorporated into the \textsc{marcs} models. Mass loss is one potential explanation, although it has been shown that the flux in both the $J$ and $K_S$ bands is relatively insensitive to the mass-loss rate \citep{Davies2021}.

To bypass the limitations of the TiO diagnostic, we derived two scaling relations using temperatures from \cite{Davies2013, Davies2015, Tabernero2018} and bridged the $T_{\rm eff,TiO}$ discrepancy. The extent of the discrepancy scales tightly with the mass-loss rate, in agreement with the increased TiO absorption for mass-losing RSGs shown by \cite{Davies2021}. We applied the scaling relation to our modeled $T_{\rm eff,TiO}$, and compared the results to evolutionary tracks. The $i$-band scaled TiO temperatures are a good match to the core helium burning stage of the \textsc{Posydon} evolutionary tracks. A metallicity and mass-loss rate dependence of the scaling relations should be established in a future study.

We have compared 33 dusty RSGs ($\textrm{m}_{3.6}-\textrm{m}_{4.5}>0.1$~mag and M$_{3.6}\le-9.0$~mag) to non-dusty RSGs and found that the dusty RSGs are on average \textit{i)} 60~K cooler, \textit{ii)} more extinct in the $V$-band by $\sim 0.6$~mag and \textit{iii)} more luminous by 0.25~dex, indicating that RSGs with circumstellar dust are generally more evolved. Their IR excess was therefore expected, as they are more likely to shed mass and form dust. Three dusty RSGs exceed the revised Humphreys-Davidson limit and have luminosities and stellar radii comparable to WOH~G64. We also found three K-type RSGs with IR excess and speculate that these are candidate Levesque-Massey variables that may have transitioned from a cooler M-type RSG back to a hotter state in recent years. Alternatively, we speculate that these may have experienced a recent, anisotropic mass ejection outside the line of sight, leaving the surface unobscured, but enriching the circumstellar environment with dust. 

Lastly, we propose that the TiO diagnostic has great potential to probe the mass-loss rates of RSGs. Following \cite{Davies2021}, who demonstrated the effects of mass loss on the RSG spectra, we tentatively argue that the existence of a complete grid of \textsc{marcs} models incorporating such effects could potentially benchmark a relation between mass loss and metallicity for RSGs. Obtaining microturbulent velocities from broadened spectral lines in high-resolution spectra (using e.g., \textsc{marcs} models) may provide empirical insights into the driving mechanism of RSG winds. We suggest that when $Z$ and $T_{\rm eff}$ are known from near-IR spectroscopy (e.g., KMOS), the TiO bands could serve as a direct and easily accessible diagnostic to derive the mass-loss rates of distant RSGs.

\begin{acknowledgements}
S.dW., A.Z.B., K.A., E.Z., E.C., G.M., and G.M.S. acknowledge funding support from the European Research Council (ERC) under the European Union’s Horizon 2020 research and innovation program (Grant agreement No. 772086. E.Z. also acknowledges support from the Hellenic Foundation for Research and Innovation (H.F.R.I.) under the ``3rd Call for H.F.R.I. Research Projects to Support Post-Doctoral Researchers'' (Project No: 7933). N.B. acknowledges support from the postdoctoral program (IPD-STEMA) of Liege University, and the Belgian federal government grant for Ukrainian
postdoctoral researchers (contract UF/2022/10). K.D.'s work was supported by NASA through the NASA Hubble Fellowship grant \#HST-HF2-51477.001 awarded by the Space Telescope Science Institute, which is operated by the Association of Universities for Research in Astronomy, Inc., for NASA, under contract NAS5-26555. Based on observations collected at the European Southern Observatory under ESO program 105.20HJ and 109.22W2. This research has made use of NASA's Astrophysics Data System. This research has made use of the SVO Filter Profile Service (\url{http://svo2.cab.inta-csic.es/theory/fps/}) supported by the Spanish MINECO through grant AYA2017-84089. \\
\textit{Spitzer:} This work is based in part on observations made with the \textit{Spitzer Space Telescope}, which is operated by the Jet Propulsion Laboratory, California Institute of Technology under a contract with NASA.\\
\textit{Gaia:} This work has made use of data from the European Space Agency (ESA) mission {\it Gaia} (\url{https://www.cosmos.esa.int/gaia}), processed by the {\it Gaia} Data Processing and Analysis Consortium (DPAC, \url{https://www.cosmos.esa.int/web/gaia/dpac/consortium}). Funding for the DPAC has been provided by national institutions, in particular, the institutions participating in the {\it Gaia} Multilateral Agreement. \\
\textit{WISE:} This publication makes use of data products from the Wide-field Infrared Survey Explorer, which is a joint project of the University of California, Los Angeles, and the Jet Propulsion Laboratory/California Institute of Technology, funded by the National Aeronautics and Space Administration. \\
\textit{Pan-STARRS1:} The Pan-STARRS1 Surveys (PS1) and the PS1 public science archive have been made possible through contributions by the Institute for Astronomy, the University of Hawaii, the Pan-STARRS Project Office, the Max-Planck Society and its participating institutes, the Max Planck Institute for Astronomy, Heidelberg and the Max Planck Institute for Extraterrestrial Physics, Garching, The Johns Hopkins University, Durham University, the University of Edinburgh, the Queen's University Belfast, the Harvard-Smithsonian Center for Astrophysics, the Las Cumbres Observatory Global Telescope Network Incorporated, the National Central University of Taiwan, the Space Telescope Science Institute, the National Aeronautics and Space Administration under Grant No. NNX08AR22G issued through the Planetary Science Division of the NASA Science Mission Directorate, the National Science Foundation Grant No. AST-1238877, the University of Maryland, Eotvos Lorand University (ELTE), the Los Alamos National Laboratory, and the Gordon and Betty Moore Foundation. \\
\textit{ATLAS:} This work includes data from the Asteroid Terrestrial-impact Last Alert System (ATLAS) project. ATLAS is primarily funded to search for near-earth asteroids through NASA grants NN12AR55G, 80NSSC18K0284, and 80NSSC18K1575; byproducts of the NEO search include images and catalogs from the survey area. The ATLAS science products have been made possible through the contributions of the University of Hawaii Institute for Astronomy, the Queen's University Belfast, the Space Telescope Science Institute, and the South African Astronomical Observatory.
\end{acknowledgements}


\bibliographystyle{aa} 
\bibliography{bib.bib} 

\begin{appendix}
\section{Supplementary figures} \label{App1}

\begin{figure*}[h]
\begin{subfigure}[t]{0.5\textwidth}
    \includegraphics[width=1\columnwidth]{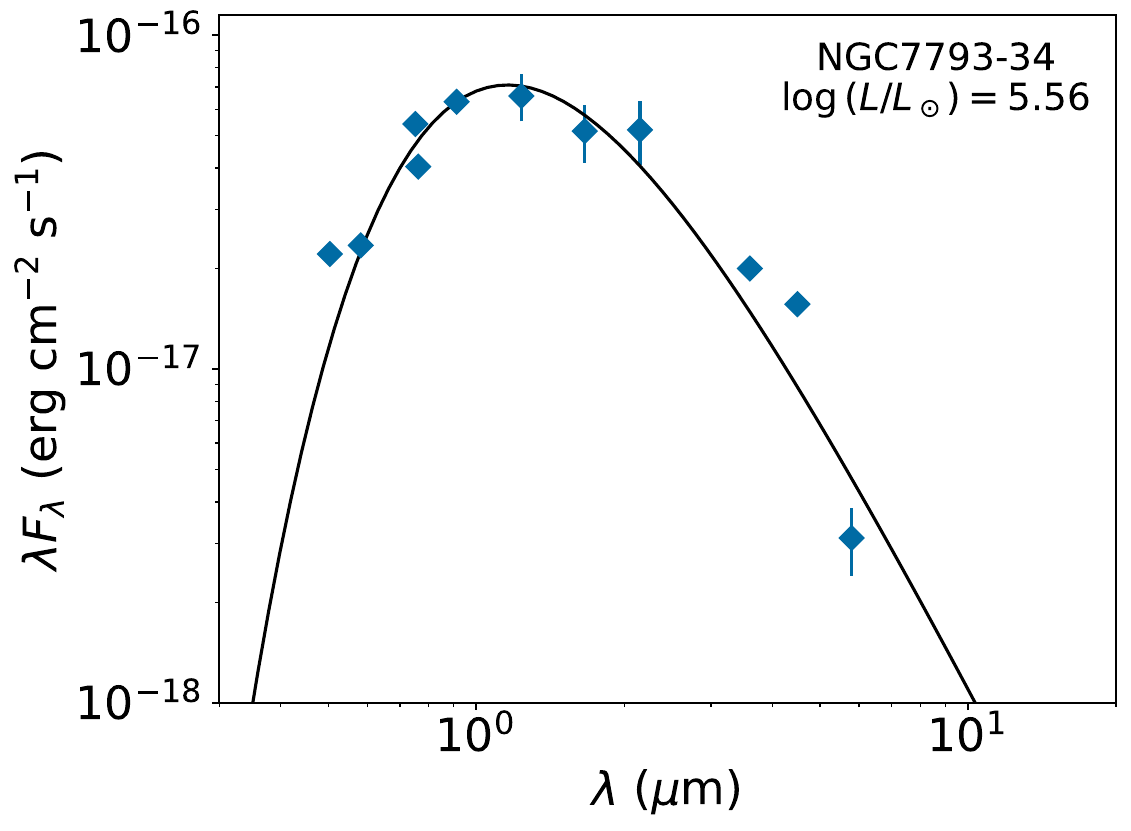}
\end{subfigure}
\begin{subfigure}[t]{0.5\textwidth}
    \includegraphics[width=1\columnwidth]{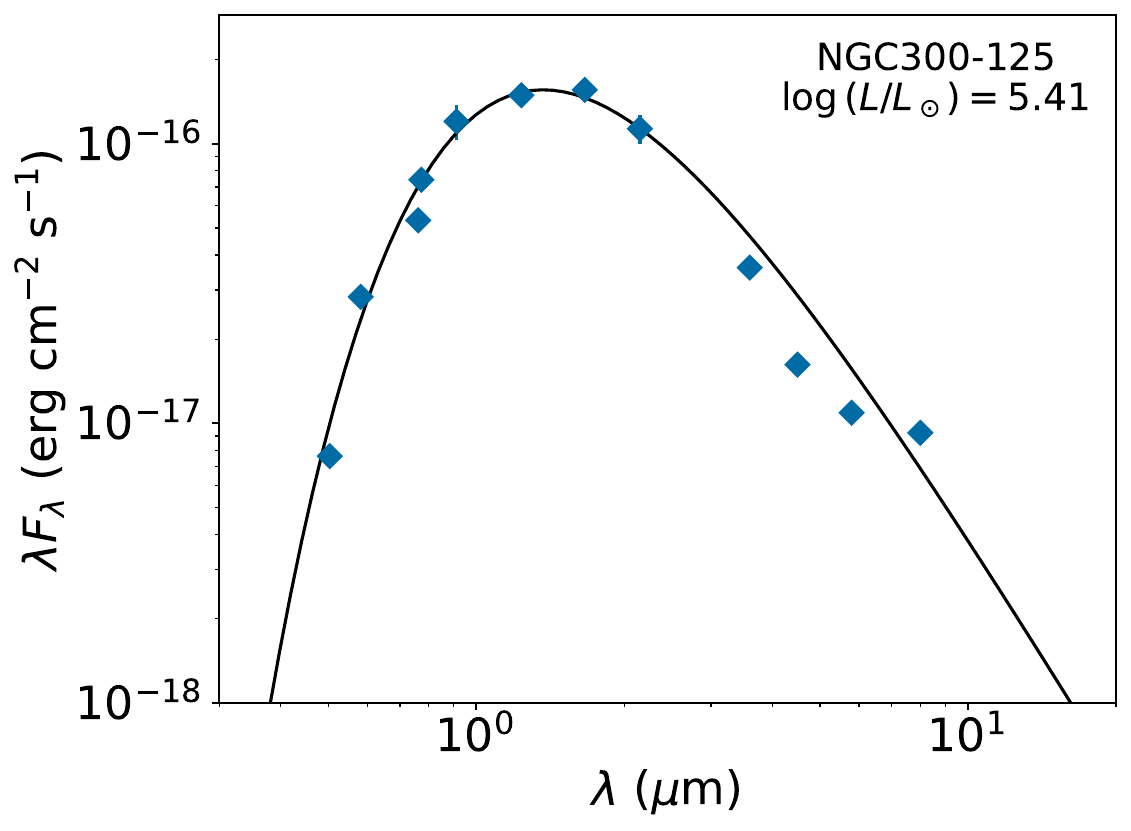}
\end{subfigure}
\begin{subfigure}[t]{0.5\textwidth}
    \includegraphics[width=1\columnwidth]{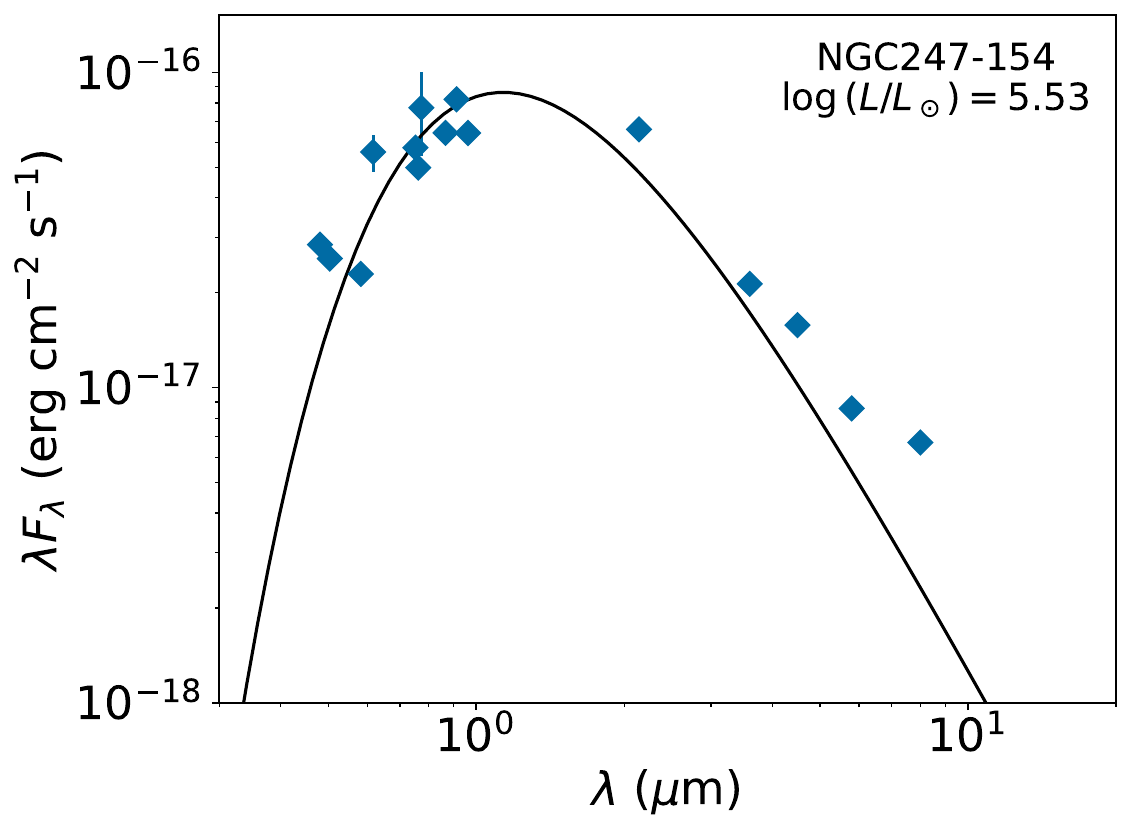}
\end{subfigure}
\begin{subfigure}[t]{0.5\textwidth}
    \includegraphics[width=1\columnwidth]{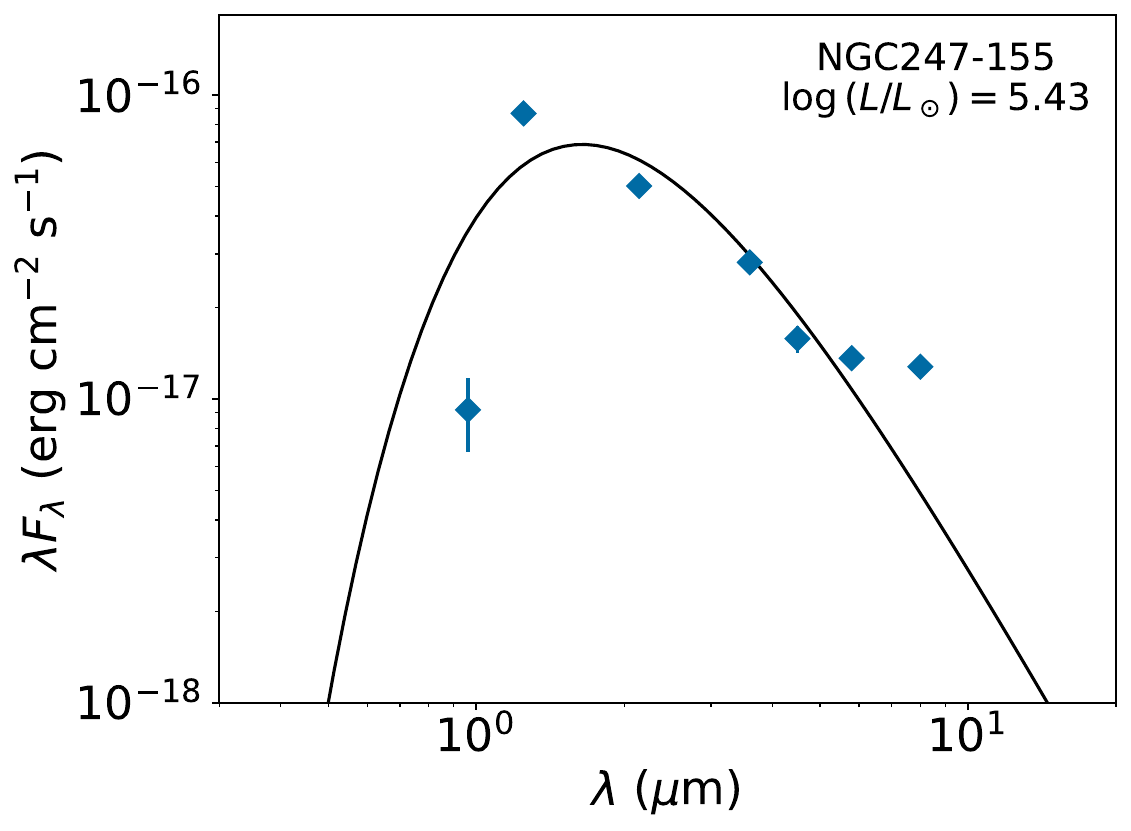}
\end{subfigure}
\begin{subfigure}[t]{0.5\textwidth}
    \includegraphics[width=1\columnwidth]{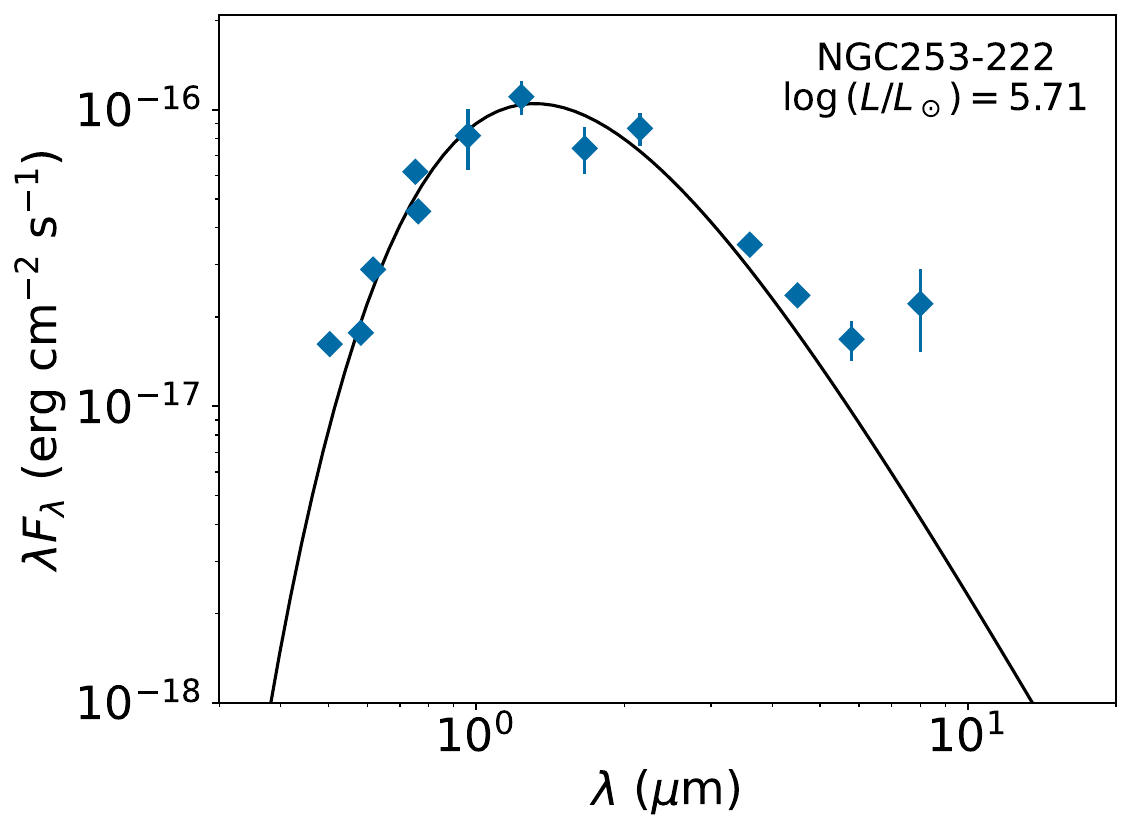}
\end{subfigure}
\begin{subfigure}[t]{0.5\textwidth}
    \includegraphics[width=1\columnwidth]{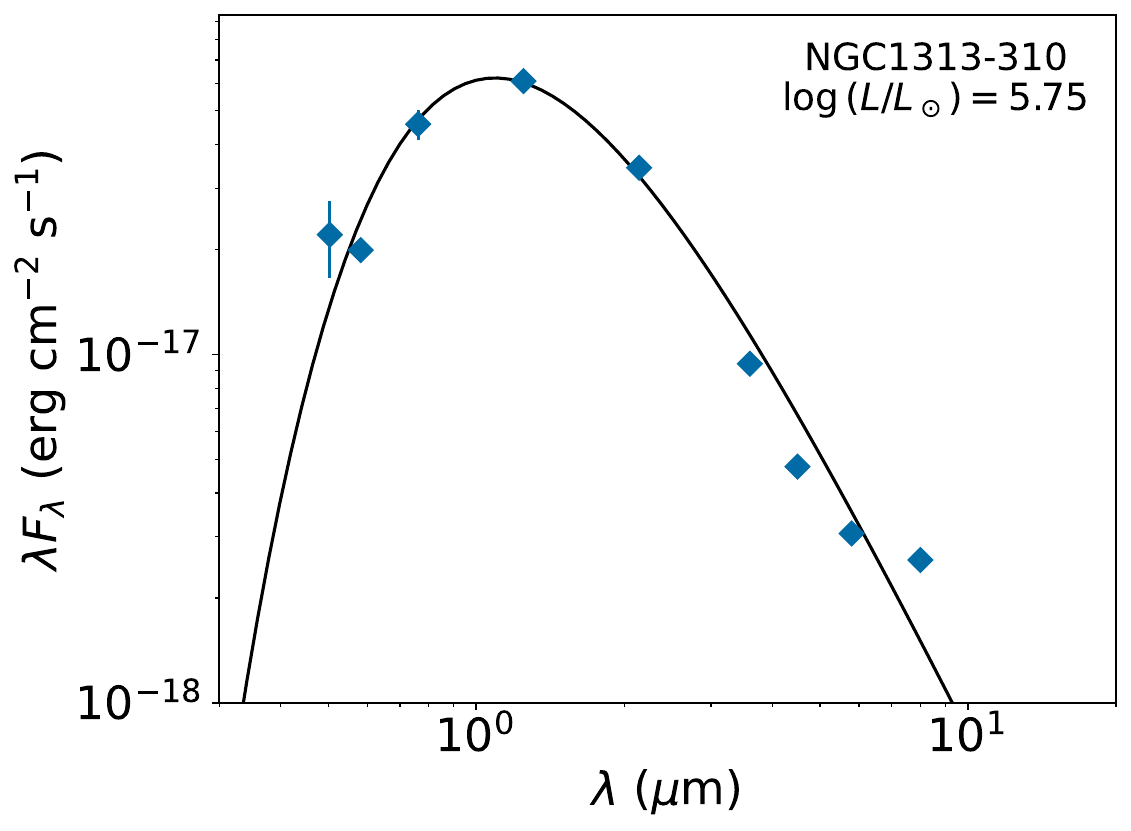}
\end{subfigure}

\caption{Blackbody fits of six extreme luminosity RSGs presented in Sects.~\ref{SecHRD} and \ref{SecDusty}. The solid line is the best-fit black-body curve to the observed SED, yielding the luminosity. Blue diamonds show the photometric observations in different bands.}
\label{fig:HRD_Lfits}
\end{figure*}

\begin{figure*}[h]
\begin{subfigure}[t]{0.5\textwidth}
\centering
    \includegraphics[width=0.92\columnwidth]{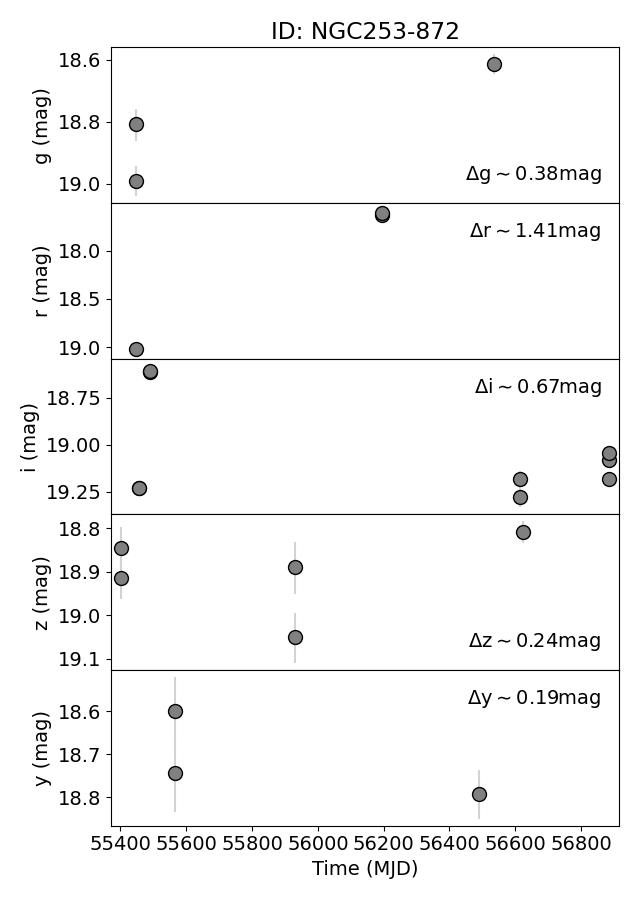}
\end{subfigure}
\begin{subfigure}[t]{0.5\textwidth}
\centering
    \includegraphics[width=0.92\columnwidth]{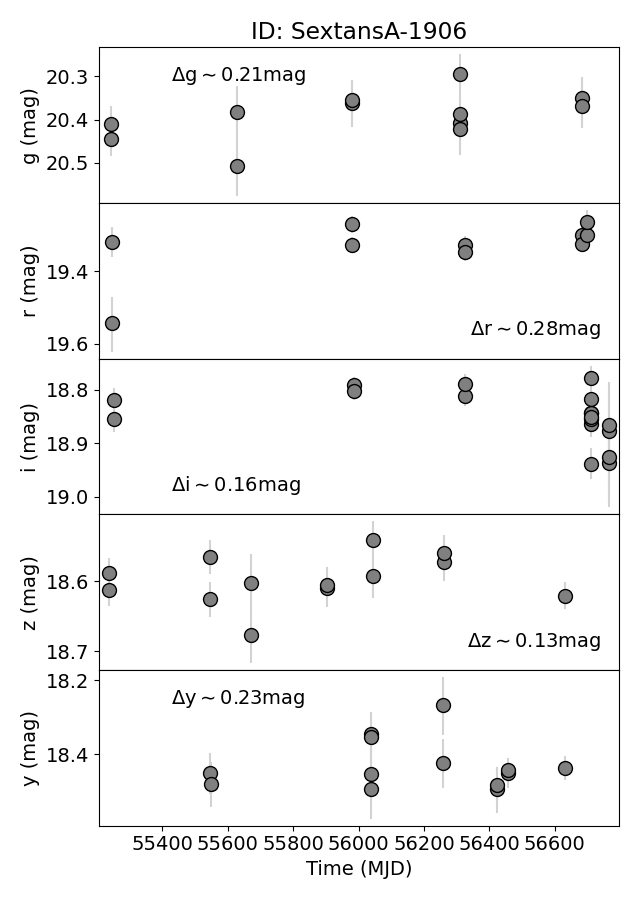}
\end{subfigure}
\begin{subfigure}[t]{0.5\textwidth}
\centering
    \includegraphics[width=0.92\columnwidth]{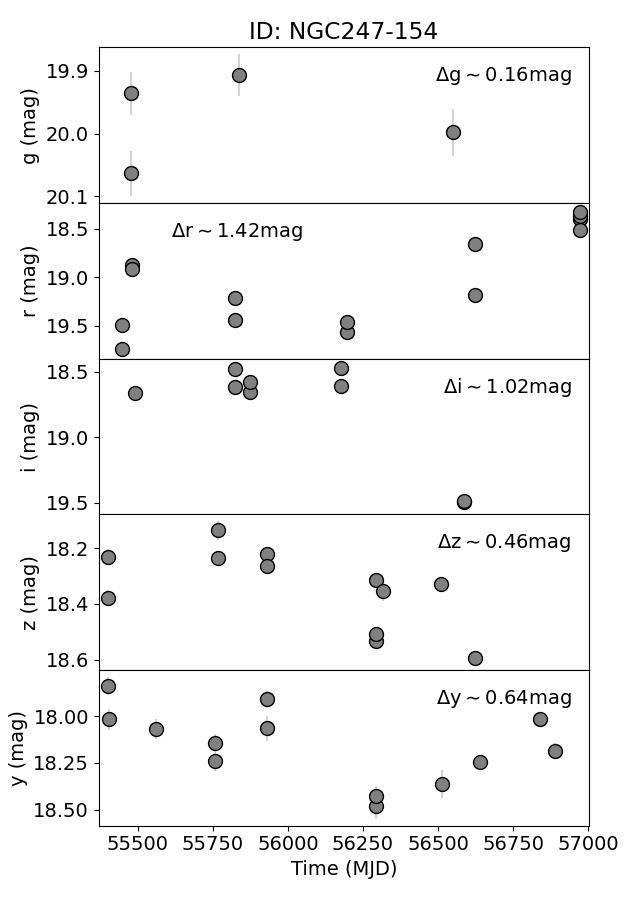}
\end{subfigure}
\begin{subfigure}[t]{0.5\textwidth}
\centering
    \includegraphics[width=0.92\columnwidth]{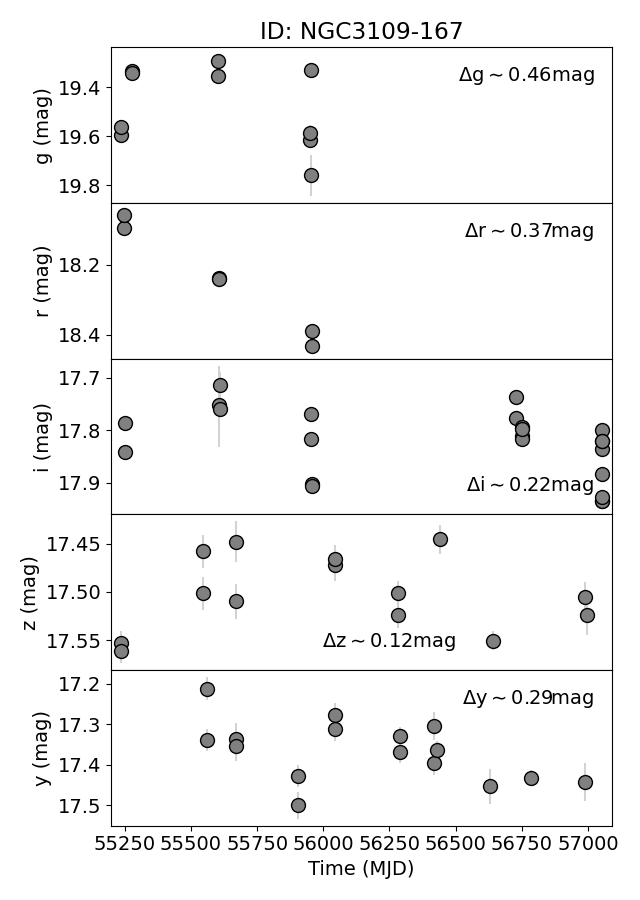}
\end{subfigure}

\caption{Complete Pan-STARRS1 $grizy$ light curves of 4 RSGs presented in Sect.~\ref{Data:LCs} (NGC3109-167, SextansA-1906, NGC247-154, NGC253-872).}
\label{fig:AllPS1}

\end{figure*}

\begin{figure*}[h]
\begin{subfigure}[t]{1\textwidth}
\centering
    \includegraphics[width=0.8\columnwidth]{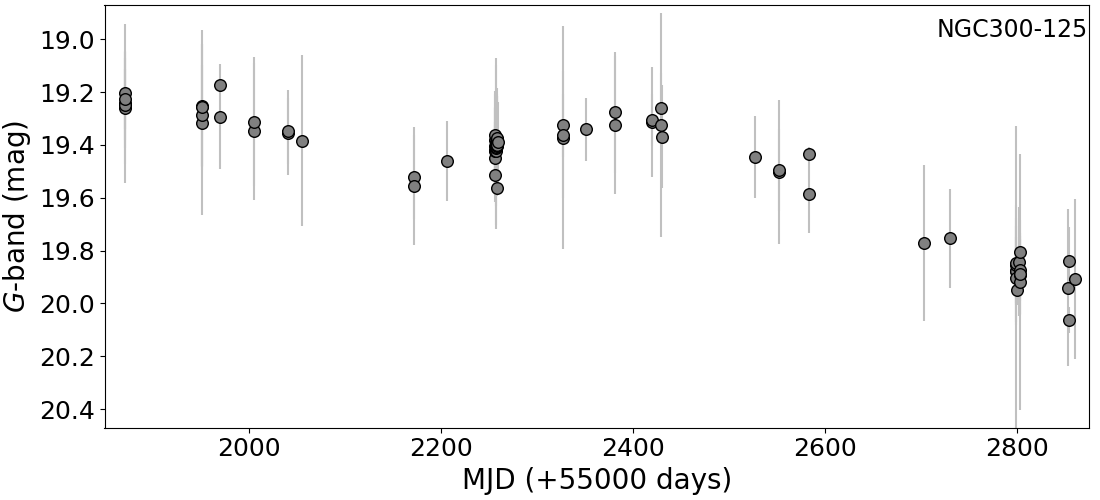}
\end{subfigure}
\begin{subfigure}[t]{1\textwidth}
\centering
    \includegraphics[width=0.8\columnwidth]{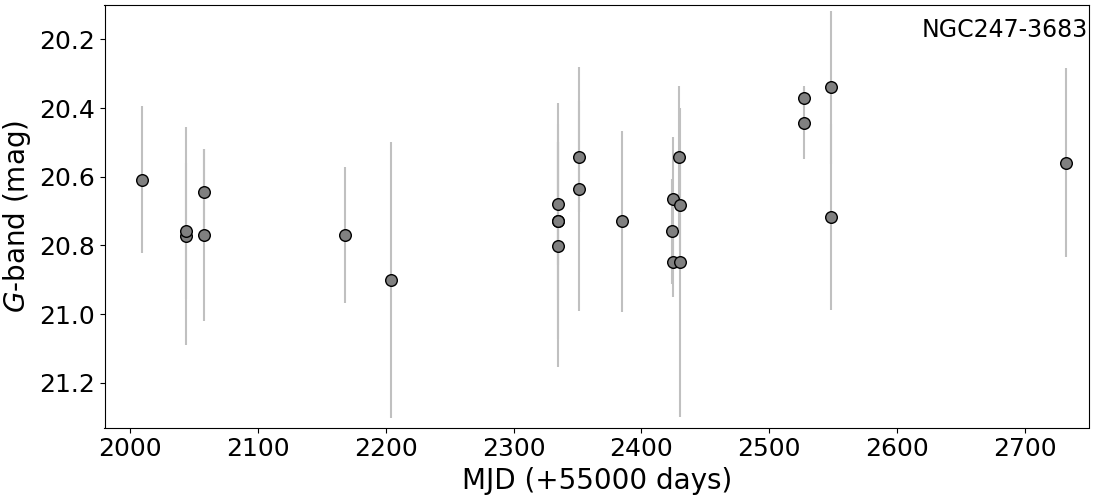}
\end{subfigure}
\begin{subfigure}[t]{1\textwidth}
\centering
    \includegraphics[width=0.8\columnwidth]{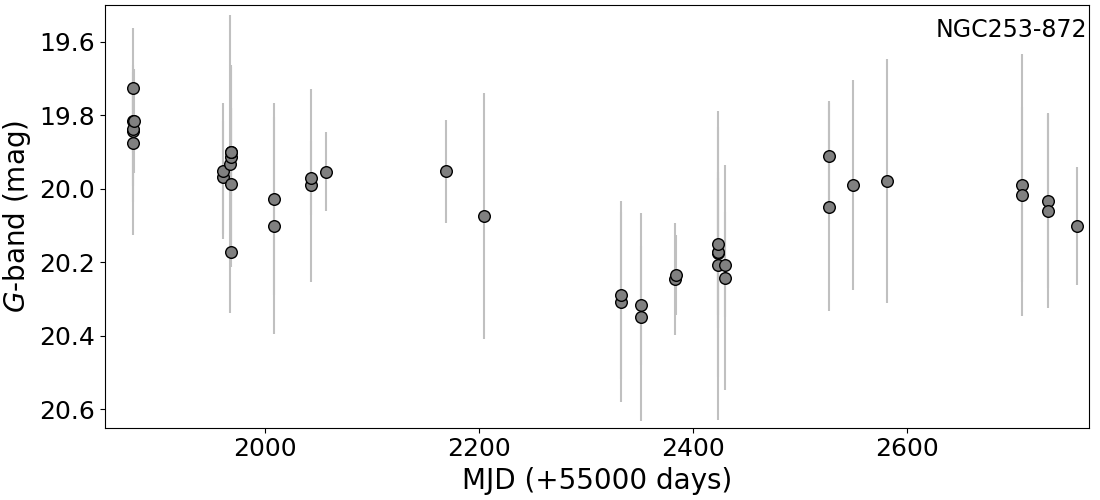}
\end{subfigure}

\caption{\textit{Gaia} light curves of NGC300-125, NGC247-3683, and NGC253-872 presented in Sect.~\ref{SecVarOpt}.}
\label{fig:AllGaia}
\end{figure*}

\begin{figure*}[h]
\begin{subfigure}[t]{0.5\textwidth}
    \includegraphics[width=0.98\columnwidth]{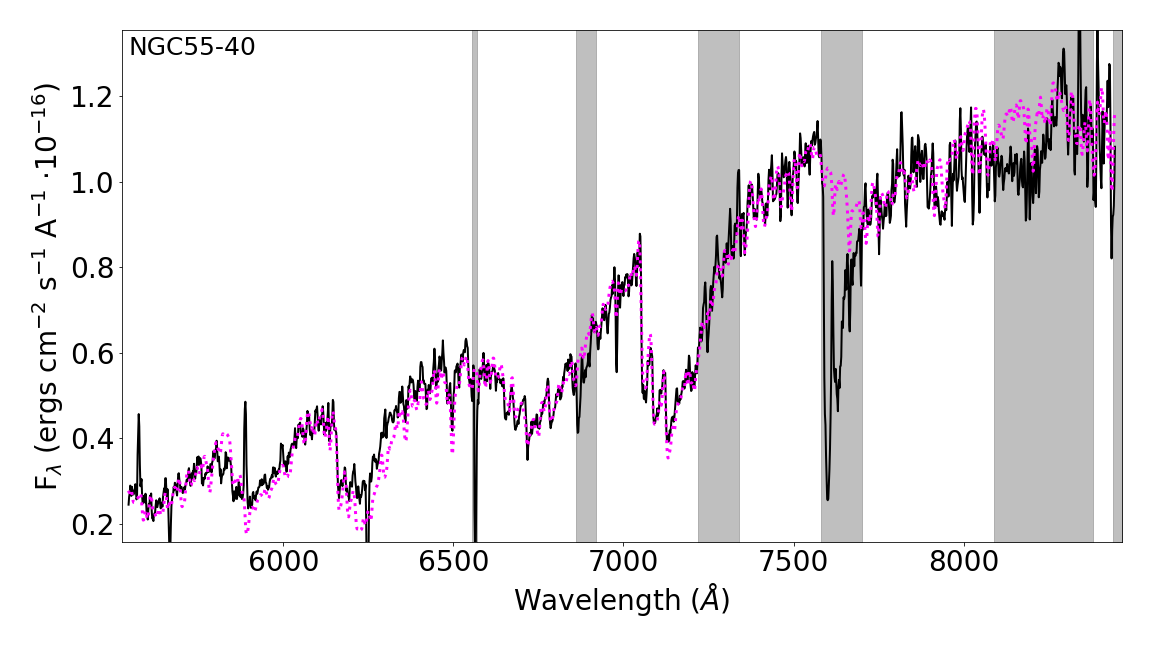}
\end{subfigure}
\begin{subfigure}[t]{0.5\textwidth}
    \includegraphics[width=0.98\columnwidth]{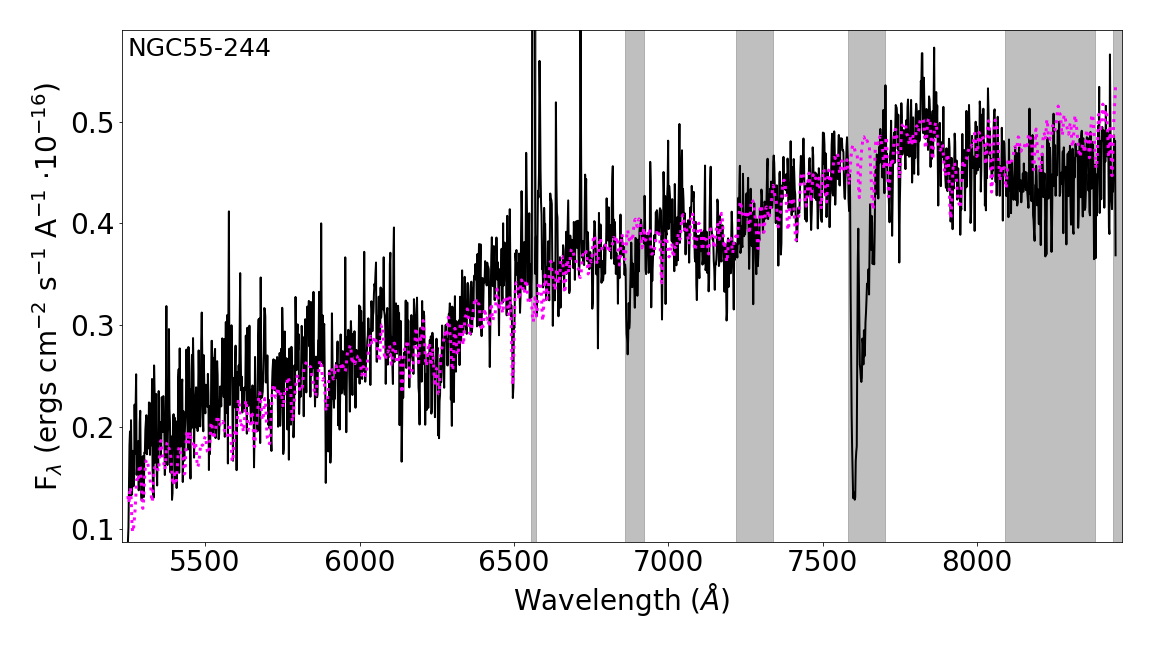}
\end{subfigure}
\begin{subfigure}[t]{0.5\textwidth}
    \includegraphics[width=0.98\columnwidth]{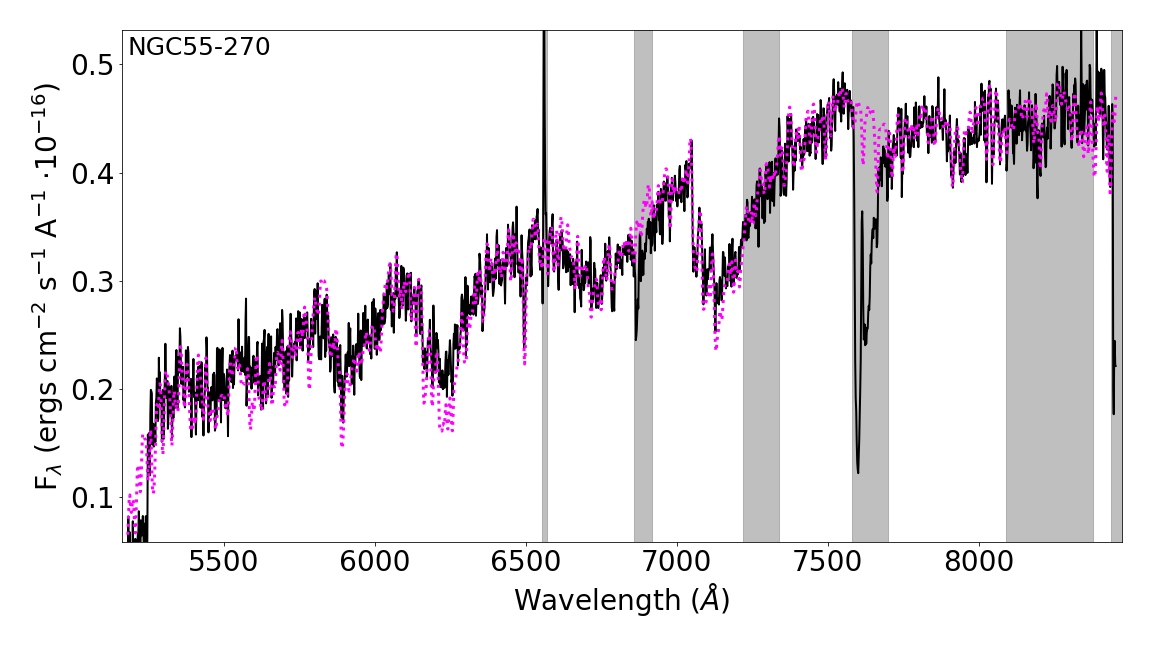}
\end{subfigure}
\begin{subfigure}[t]{0.5\textwidth}
    \includegraphics[width=0.98\columnwidth]{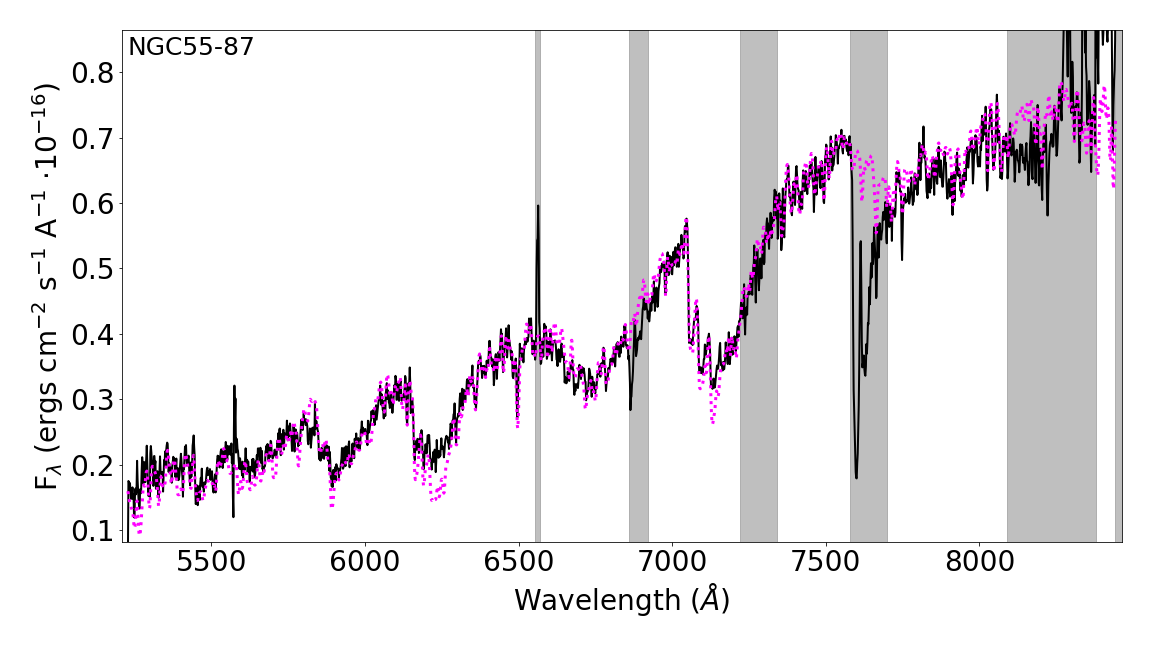}
\end{subfigure}
\begin{subfigure}[t]{0.5\textwidth}
    \includegraphics[width=0.98\columnwidth]{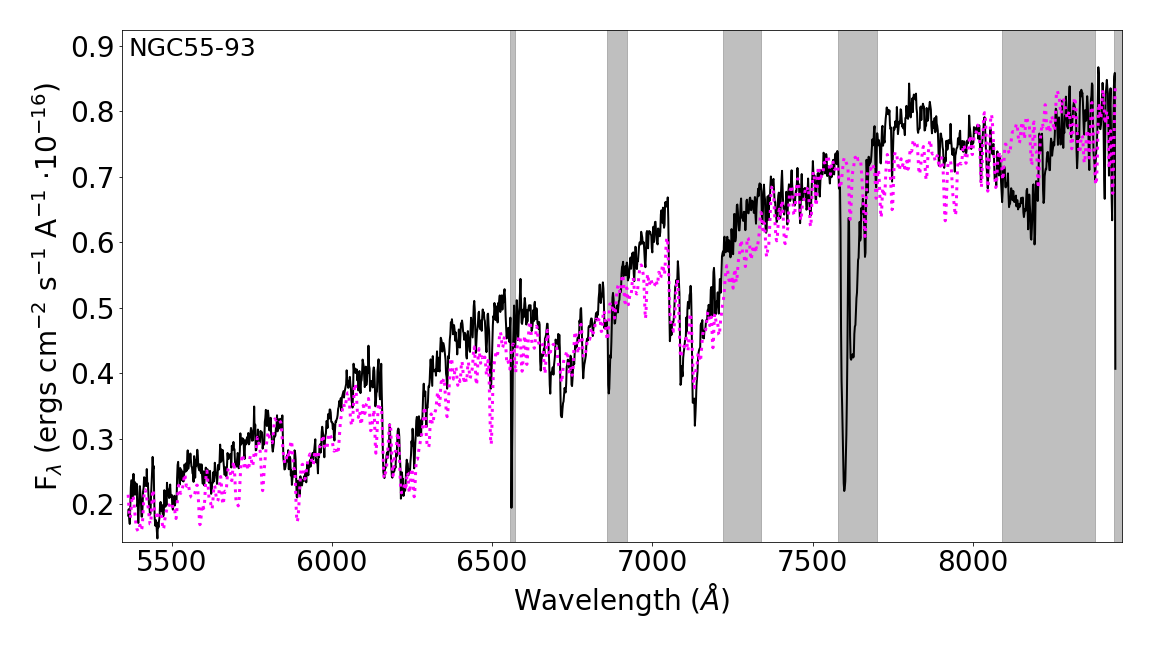}
\end{subfigure}
\begin{subfigure}[t]{0.5\textwidth}
    \includegraphics[width=0.98\columnwidth]{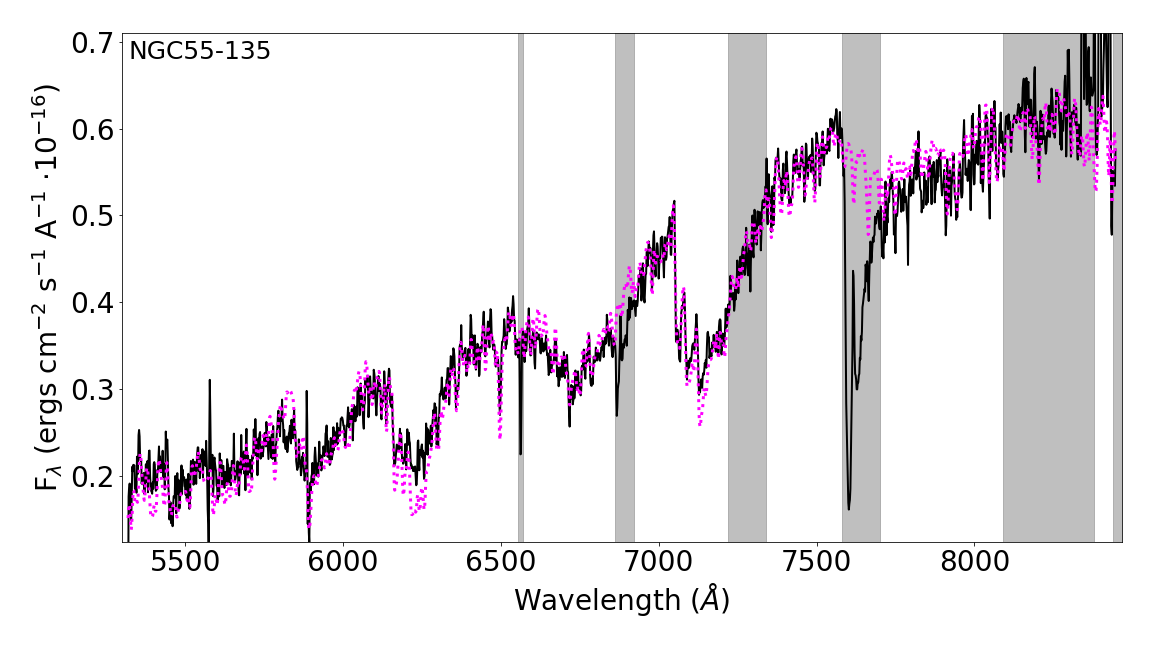}
\end{subfigure}
\begin{subfigure}[t]{0.5\textwidth}
    \includegraphics[width=0.98\columnwidth]{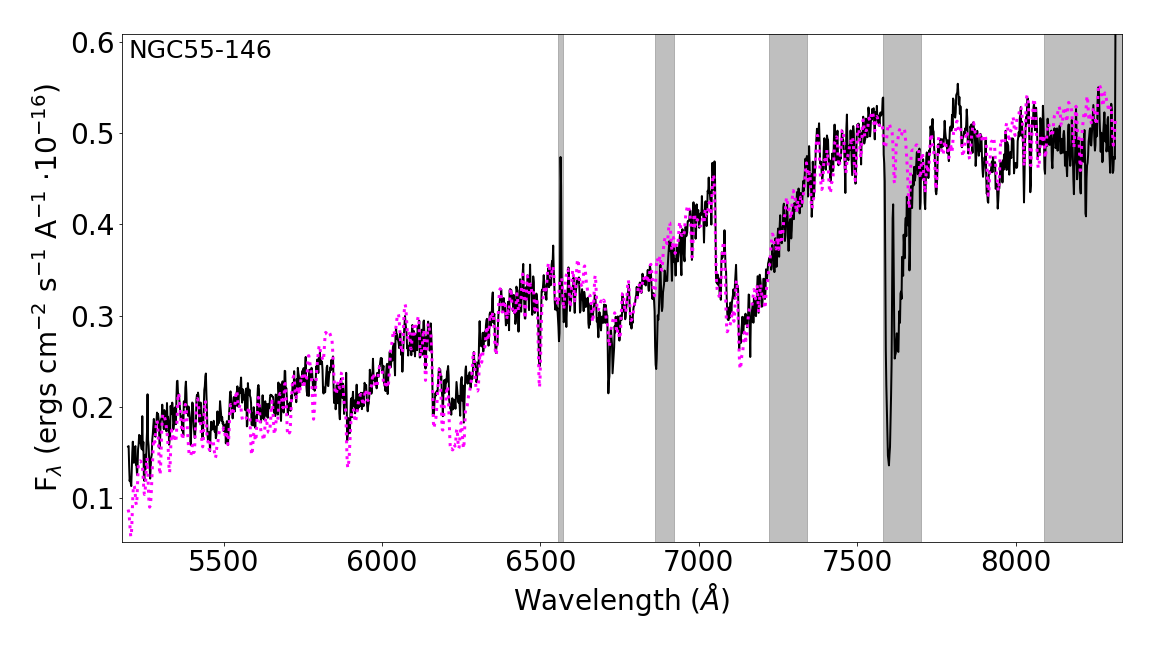}
\end{subfigure}
\begin{subfigure}[t]{0.5\textwidth}
    \includegraphics[width=0.98\columnwidth]{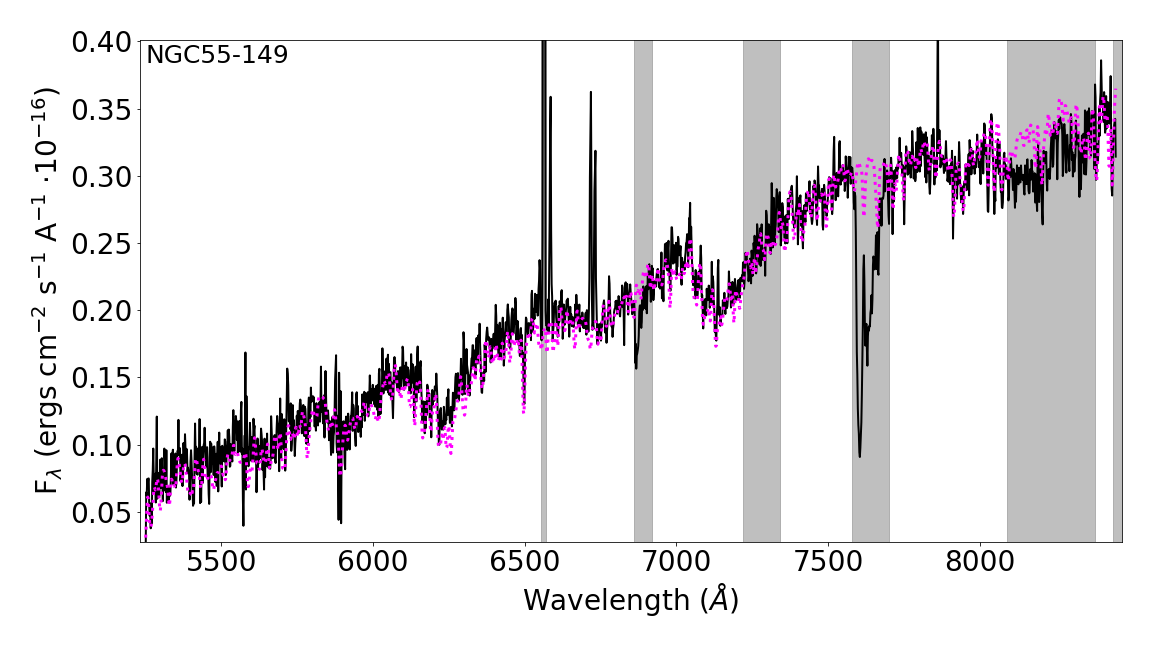}
\end{subfigure}
\begin{subfigure}[t]{0.5\textwidth}
    \includegraphics[width=0.98\columnwidth]{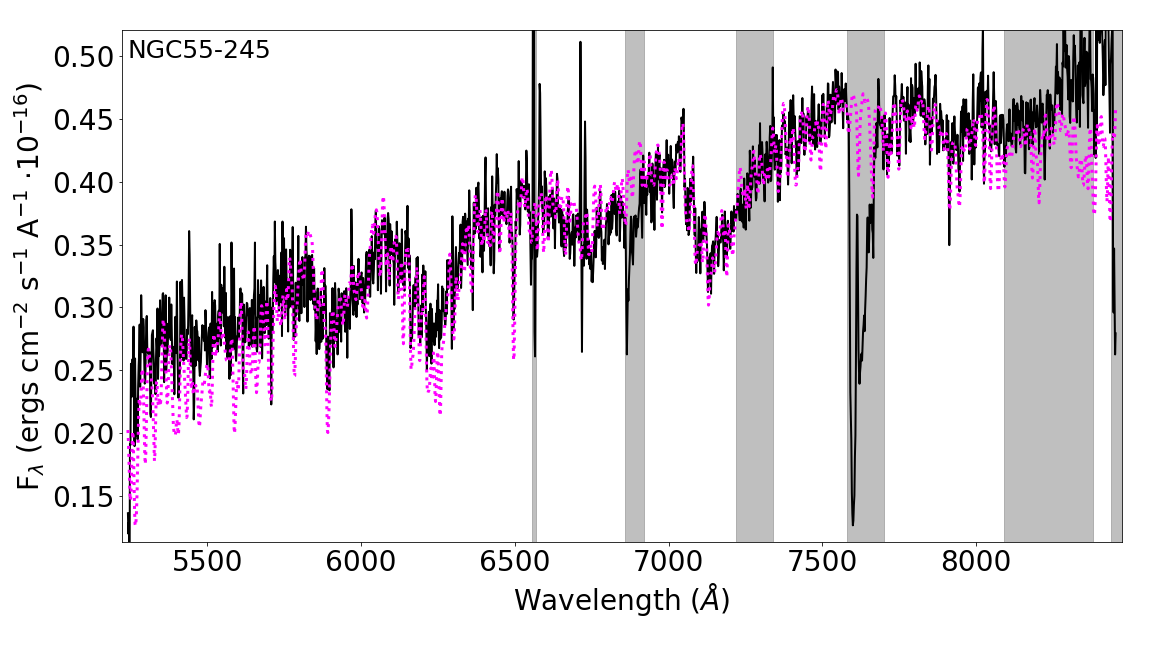}
\end{subfigure}
\begin{subfigure}[t]{0.5\textwidth}
    \includegraphics[width=0.98\columnwidth]{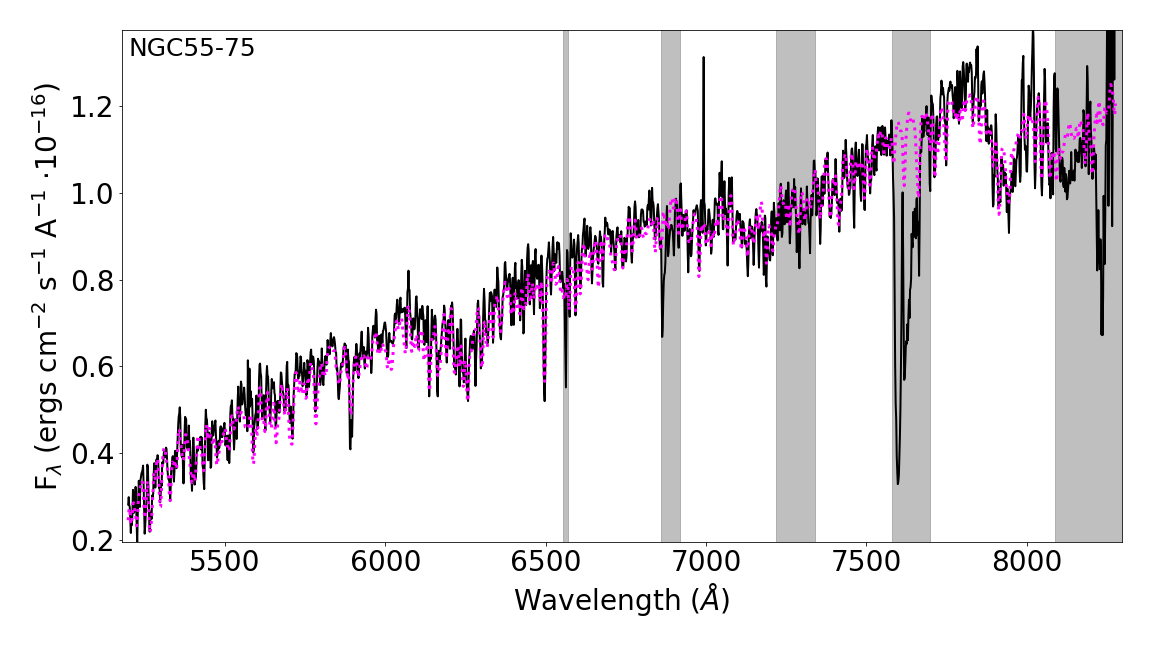}
\end{subfigure}

\caption{Best-fit \textsc{marcs} models to all RSG spectra.}
\label{fig:AllMarcs}
\end{figure*}
\end{appendix}
\end{document}